\documentclass[compsoc, conference, a4paper, 10pt, times]{IEEEtran}

\ifCLASSOPTIONcompsoc
  \usepackage[nocompress]{cite}
\else
  \usepackage{cite}
\fi

\usepackage[numbers,sort&compress]{natbib}

\usepackage{hyperref}
\usepackage{bookmark}
\usepackage[normalem]{ulem}
\usepackage{xcolor} 
\usepackage{color}
\usepackage{booktabs} %
\usepackage{listings} %
\usepackage{subcaption}
\usepackage{amssymb}
\usepackage{amsmath}
\usepackage{cleveref}
\usepackage{graphicx}
\usepackage{scrextend}
\usepackage{pifont}
\usepackage[export]{adjustbox}
\usepackage{numprint} %
\usepackage{arydshln} %
\usepackage{multirow}
\usepackage{flushend}
\usepackage{comment}
\usepackage{tabularx}
\usepackage{makecell}
\usepackage{adjustbox}
\usepackage[colorinlistoftodos]{todonotes}
\usepackage{boldline}
\usepackage{pifont} %

\usepackage{url}

\usepackage{breakurl}

\usepackage{enumitem} %
\npdecimalsign{.}
\npthousandsep{,}

\definecolor{lightgray}{rgb}{0.95, 0.95, 0.95}
\definecolor{darkgray}{rgb}{0.4, 0.4, 0.4}
\definecolor{blueCode}{rgb}{0, 0, 0.93} %
\definecolor{greenCode}{rgb}{0, 0.6, 0} %

\Crefformat{app}{Appendix #2#1#3}
\Crefformat{table}{Table #2#1#3}

\newcommand{\empirical}[1]{\setlength{\fboxsep}{1pt}\fbox{#1}}

\newcommand{\hide}[1]{}

\newcommand{\pagesBudget}[1]{}
\renewcommand{\empirical}[1]{#1}

\newcommand\objst{\textit{Object storage}}
\newcommand\webhSame{\textit{Website hosting}}
\newcommand\webhDiff{\textit{CDN}}
\newcommand\uncat{\textit{Undetermined hosting type}}
\newcommand\toolName{\textit{Grape}}
\newcommand\pdfCheckPipeline{\textit{PDF Status Check}}
\newcommand\pPDF{clickbait PDF}
\newcommand\PPDF{Clickbait PDF}

\newcommand\dsVT{\textit{Seed DS}}
\newcommand\dsFromUrl{\textit{Main DS}}

\newcommand\identifiedVulnerableComponents{eight}

\newcommand\samplesInTraining{\numprint{23098}}
\newcommand\campaignsInTraining{\numprint{47}}

\newcommand\fqdnEntireStudy{\numprint{177835}}

\newcommand\durationEntireStudy{\numprint{17}}
\newcommand\durationPrelimStudy{three}
\newcommand\durationStudy{\numprint{13}}
\newcommand\totUniquePDFs{\numprint{609576}}

\newcommand\percPDFsCisco{\numprint{29}}
\newcommand\percPDFsInquest{\numprint{69}}

\newcommand\totPDFsVTPrelimStudy{\numprint{105598}}
\newcommand\totSeoPDFsVTPrelimStudy{\numprint{66614}}
\newcommand\totPDFsVTStudy{\numprint{503978}}
\newcommand\totSeoPDFsVTStudy{\numprint{384601}}

\newcommand\totUriPrelimStudyVTSeoUris{\numprint{1350201}}
\newcommand\totPDFsFromUrl{\numprint{2710959}}
\newcommand\totMonitoredLinksFromUrl{\numprint{4648939}}
\newcommand\percNotSharedFromUrlPartners{\numprint{93}}

\newcommand\allUniqueURLsObjst{\numprint{166356}}
\newcommand\allUniqueURLsWebhSame{\numprint{853514}}
\newcommand\allUniqueURLsWebhDiff{\numprint{595385}}
\newcommand\allUniqueURLsUncat{\numprint{4126172}}

\newcommand\amtWebhSameProvs{\numprint{20}}

\newcommand\nManuallyAdded{five}
\newcommand\totHostingProviders{\numprint{54}}
\newcommand\nSimilarWeb{\numprint{23}}

\newcommand\percTldsWCategoryData{\numprint{10}}

\newcommand\catOne{Education}
\newcommand\catOneVolume{\numprint{204679}}
\newcommand\catTwo{Graphics Multimedia and Web Design}
\newcommand\catTwoVolume{\numprint{129954}}
\newcommand\catThree{Computers Electronics and Technology}
\newcommand\catThreeVolume{\numprint{116090}}
\newcommand\catFour{Web Hosting and Domain Names}
\newcommand\catFourVolume{\numprint{99165}}
\newcommand\catFive{Sports}
\newcommand\catFiveVolume{\numprint{96612}}
\newcommand\othCatVolume{\numprint{289680}}
\newcommand\noCatVolume{\numprint{2095555}}

\newcommand\campsOnlyUnkn{Two}
\newcommand\campsUnknGtNinetyEight{Two}

\newcommand\nCampaigns{ten}
\newcommand\percTorrentPDFsUncat{\numprint{91.7}}

\newcommand\provsObservedOnline{\numprint{38}}
\newcommand\provsNotOnline{\numprint{16}}

\newcommand\avgPdfLifetimeMonths{five}
\newcommand\avgExtendedLifetimeMonths{nine}
\newcommand\tldsMaxExtendedLifetime{\numprint{1818}}
\newcommand\hostTypeLongestPDFLifetime{\objst{}}
\newcommand\avgHighestPdfLifetimeMonths{six}

\newcommand\uriPathPattern{\numprint{119662}}
\newcommand\netlocPathPattern{\numprint{1016}}

\newcommand\urisScannedTot{\numprint{159403}} %
\newcommand\awsBucketsScanned{\numprint{1776}}
\newcommand\awsExistingBuckets{\numprint{243}}

\newcommand\awsBucketsACL{\numprint{67}}

\newcommand\percAwsBucketsFC{\numprint{21}}

\newcommand\percAwsBucketsReadACP{\numprint{51}}

\newcommand\percAwsBucketsW{\numprint{28}}

\newcommand\percBucketsObservedACL{\numprint{27}}

\newcommand\nDomainsCfInStudy{\numprint{85582}}
\newcommand\nDomainsCfInStudyNoWeebly{\numprint{31724}}

\newcommand\percCompFrwOverTot{\numprint{29}}
\newcommand\nSwObserved{\numprint{299}}

\newcommand\percDomainsObservationsUnkn{\numprint{96}}

\newcommand\domsOutdatedComponents{\numprint{12927}}

\newcommand\percOutdatedWebhSame{\numprint{26}}

\newcommand\percNoIndicatorDomainsWeebly{\numprint{90}}

\newcommand\componentsWVersion{\numprint{115}}

\newcommand\vulnComponents{\numprint{26}}

\newcommand\nVulnsUfu{ten}
\newcommand\ufuComponents{five}
\newcommand\ufuVulnerableDoms{\numprint{225}}

\newcommand\fqdnsAnalyzedForVulns{\numprint{11815}} %

\newcommand\nGuessedComponents{eight}
\newcommand\formcraftPathGcp{\numprint{621}}
\newcommand\webformPathGcp{\numprint{482}}
\newcommand\kcfinderPathGcp{\numprint{799}}
\newcommand\ckfinderPathGcp{\numprint{2436}}
\newcommand\fckeditorPathGcp{\numprint{232}}
\newcommand\ckeditorPathGcp{\numprint{88}}
\newcommand\SLiMSeLearnPathGcp{\numprint{1018}}
\newcommand\SLiMSeLearnCpScan{\numprint{396}}
\newcommand\vulnSLiMSCpScan{\numprint{73}}
\newcommand\vulnELearnCpScan{\numprint{38}}

\newcommand\recaptchaWIoC{\numprint{43}}
\newcommand\robloxWIoC{\numprint{52}}

\newcommand\cpEndpoints{\numprint{59}}

\newcommand\domainsObservationsCpScan{\numprint{9800}}

\newcommand\percVulnCpScanner{\numprint{55}}

\newcommand\ckfinderCpScan{\numprint{4396}}
\newcommand\vulnCkfinderCpScan{\numprint{100}}
\newcommand\ckeditorCpScan{\numprint{4840}}
\newcommand\vulnCkeditorCpScan{\numprint{91}}
\newcommand\fckeditorCpScan{\numprint{4933}}
\newcommand\vulnFckeditorCpScan{0}
\newcommand\kcfinderCpScan{\numprint{262}}
\newcommand\vulnKcfinderCpScan{\numprint{100}}

\newcommand\urlsFromIoCBuckets{\numprint{4191}}
\newcommand\urlsFromIoCUncatWebhsame{\numprint{1251059}}
\newcommand\nUrlsAmzSetupPhase{\numprint{49065}}

\newcommand\lengthGSBDataCollection{\numprint{17}}

\newcommand\percBlockedOverSubm{\numprint{0.4}}
\newcommand\nBlocklGsbNetloc{\numprint{451}}
\newcommand\avgBlocklGsbNetloc{\numprint{41}}

\newcommand\percOnlineNowBlockl{\numprint{29}}

\newcommand\minAvgPDFScore{one}
\newcommand\hostTypeMinAvgScore{\objst{}}
\newcommand\maxAvgPDFScore{five}
\newcommand\hostTypeMaxAvgScore{\webhSame{}}

\newcommand\nPDFsNotified{\numprint{799930}}
\newcommand\dimensionTestGrp{\numprint{8843}}
\newcommand\dimensionControlGrp{\numprint{8842}}
\newcommand\totSelectedContacts{\numprint{1545}}
\newcommand\successfullySentFirstNot{\numprint{1522}}
\newcommand\domainsNotReported{\numprint{19}}

\newcommand\providersNeverReached{\numprint{257}}

\newcommand\nFalsePos{\numprint{9}}
\newcommand\nOurAnswers{\numprint{124}}

\newcommand\totContactsForIPs{\numprint{32302}}
\newcommand\totFetchedIPs{\numprint{12043}}
\newcommand\totSynthDomains{\numprint{153}}

\newcommand\fhashFetchedVT{\numprint{111787}}
\newcommand\fhashKnownVT{\numprint{57042}}
\newcommand\percKnownAfterNotif{1}
\newcommand\submittedForms{\numprint{25}}

\newcommand\percProvPartialFix{\numprint{17}}

\newcommand\provCleanEmBodyOverall{\numprint{104}}
\newcommand\percProvCleanEmBodyOverall{\numprint{7}}

\newcommand\nProvsReExpl{\numprint{154}}
\newcommand\nProvsDeepClean{\numprint{319}}
\newcommand\percProvsReExpl{\numprint{10}}
\newcommand\percProvsDeepClean{\numprint{21}}

\begin{document}

\title{Uncovering the Role of Support Infrastructure in \PPDF{} Campaigns}

\author{
	\IEEEauthorblockN{Giada Stivala\IEEEauthorrefmark{1},
		Gianluca De Stefano\IEEEauthorrefmark{1}, Andrea Mengascini\IEEEauthorrefmark{1}, Mariano Graziano\IEEEauthorrefmark{2}, Giancarlo Pellegrino\IEEEauthorrefmark{1}}
	\IEEEauthorblockA{\IEEEauthorrefmark{1}
		CISPA Helmholtz Center for Information Security\\
		\{giada.stivala, gianluca.de-stefano, andrea.mengascini, pellegrino\}@cispa.de}
	\IEEEauthorblockA{\IEEEauthorrefmark{2}Cisco Talos\\
	graziano.mariano@gmail.com}
}

\maketitle

\begin{abstract} %
\PPDF s, an entry point for multiple Web attacks, are distributed via SEO poisoning and rank high in search results due to being massively uploaded on abused or compromised websites.
The central role of these hosts in the distribution of \pPDF s remains understudied, and it is unclear whether attackers differentiate the types of hosting for PDF uploads, how long they rely on hosts, and how affected parties respond to abuse.

To address this, we conducted real-time analyses on hosts, collecting data on \empirical{\totMonitoredLinksFromUrl{}} \pPDF s served by \empirical{\fqdnEntireStudy{}} hosts over \empirical{\durationEntireStudy{}} months. Our results revealed a diverse infrastructure, with hosts falling into three main hosting types. We also identified at scale the presence of  \empirical{\identifiedVulnerableComponents{}} software components which facilitate file uploads and which are likely exploited for \pPDF{} distribution.
We contact affected parties to report the misuse of their resources via a large-scale vulnerability notification. While we observed some effectiveness in terms of number of cleaned-up PDFs following the notification, long-term improvement in this infrastructure remained insignificant. 
This finding raises questions about the hosting providers' role in combating abuse and the actual impact of vulnerability notifications.
\end{abstract}

\section{Introduction} \pagesBudget{1}
Phishing and spam have been known for decades and, nonetheless, they keep being a profitable option for cybercriminals~\cite{ic3report}. While most known for spreading via e-mail~\cite{stone2011underground}, mediums for these attacks have evolved in time, encompassing a variety of means, such as SMS, phone calls, social media platforms~\cite{lee2012warningbird, stringhini2010detecting}, and the more recent \pPDF s~\cite{stivala23from}.
\PPDF s are PDF documents whose first page contains a visual bait embedding a link to a Web attack such as phishing attacks~\cite{stivala23from}, malware download~\cite{stivala23from}, and malicious browser extension download~\cite{paloalto_pdftrend}. They are distributed via search engine (SE) poisoning attacks, ranking high as these files are hosted in benign servers and cross-reference one another, increasing their page rank.
Only recently the research community has started looking into this new threat, focusing on visual baits~\cite{stivala23from}, type of Web attack~\cite{paloalto_pdftrend, microsoftPhishingAttachments}, and volumetric features~\cite{paloalto_pdftrend, stivala23from}, but neglecting the role played by the supporting infrastructure in these attacks.

The effectiveness of \pPDF{} attacks relies on massive daily uploads of cross-linked PDFs in benign servers~\cite{stivala23from}.
Studying the supporting infrastructure has been a critical aspect when analyzing other similar threats, such as drive-by download~\cite{nappa2013driving}, phishing pages~\cite{oest2018inside}, spam~\cite{liao2016characterizing}, or when looking at server compromise~\cite{catakoglu2016automatic, john2011deseo} and their role in the attacks~\cite{oest2018inside}.
Despite previous research efforts, there are still gaps in our understanding of the supporting infrastructure. 
These findings do not directly apply to \pPDF s, as they focus on threats with different core features or examine the infrastructure with limited scope.
Firstly, it remains unclear which and how many types of hosting are utilized to serve \pPDF s. 
Prior works, when considering the type of hosting in their analyses, have used it as a pre-filtering criterion (e.g., only Cloud Storage~\cite{liao2016characterizing}) or focused on a predefined list of domains~\cite{liao2016lurking}.
Another key feature of this threat is the extensive volume of PDFs online on infrastructure hosts for prolonged periods, ensuring their presence in poisoned search results~\cite{stivala23from}. 
This temporal aspect differs among malicious actions where, for example, phishing pages stay online for 1-2 days~\cite{oest2020sunrise, oest2018inside}, malware components for maximum 5.5 days~\cite{nappa2013driving} and scripts for SEO pages for a maximum of 30 days~\cite{liao2016lurking}.
The duration of the uptime of malicious resources is an attack-specific characteristic, and does not transfer straightforwardly to the threat posed by \pPDF s.
In fact, we lack knowledge about the duration of PDFs remaining online on these hosts and how website owners and hosting providers respond to such abuse of their resources.
Finally, it is unclear whether attackers rely on compromised websites as support hosts and which software component they exploited for the upload of \pPDF s.
While previous works showed that it is possible to find compromised websites by starting from vulnerable components~\cite{john2010searching, catakoglu2016automatic}, we take the opposite approach and empirically enumerate the different exploits attackers might have used to upload \pPDF s on the hosts supporting the attack.

This paper sheds light on and provides a comprehensive description of the infrastructure behind \pPDF{} attacks. Employing a data-driven approach, we conduct a large-scale study of website abuse aimed at distributing \pPDF s. 
Starting from a feed of real-world PDFs, we identify \pPDF s in this feed and leverage their cross-link structure to identify a large portion of the supporting infrastructure, which we further study with an array of specific analyses.
Our measurements and observations quantify the volume of hosts involved in this phenomenon and reveal the impact on various hosting types. Additionally, we identify \empirical{three} affected hosting types (\objst{}, \webhSame{}, and \webhDiff{}), demonstrating that the \pPDF s threat spans across multiple hosting categories.
Additionally, this paper investigates the factors exploited by attackers to gain access to these hosts.
We collect metadata on software components running at the origins in our dataset and possible indicators of compromise with analyses tailored to the specific type of hosting. 
Our findings reveal a fragmented picture, where attackers leverage characteristics specific to the type of hosting, or provider, to gain access to hosting space. For example, we identified \empirical{\identifiedVulnerableComponents{}} software components that facilitate file uploads, along with \empirical{\domsOutdatedComponents{}} origins running outdated software.

This paper also investigates ways to help mitigate the distribution of \pPDF s, whose threat is ongoing since 2020~\cite{stivala23from}.
We identify hosting providers as entities affected by this malicious action, whose resources are abused to serve \pPDF s to victim users, and reach out to them to seek their cooperation in fighting this abuse.
We undertake a vulnerability notification procedure to limit the distribution of \pPDF s, raise awareness of this threat and collect any feedback from affected parties.
Our observations show statistically significant results in the cleanup of phishing PDFs, with an overall positive feedback from the notified parties.
Worryingly, the benefits of this action appear not to last long. Notified websites serve previously unseen \pPDF s in \empirical{97\%} of the cases, indicating that most providers do not deal with the originating causes allowing file uploads.

In summary, our paper makes the following contributions:
\begin{itemize}
	\item We present a comprehensive picture of the infrastructure supporting \pPDF{} attacks, regardless of the hosting type or provider, and empirically define the volume and the duration of the abuse.
	\item We identify \empirical{three} types of hosting the websites in our dataset belong to via a systematic methodology and a thorough manual validation.
	\item We identify outdated and vulnerable software components, likely connected to host exploitation and to file upload. 
	\item We help mitigate the spread of \pPDF s by informing \empirical{\totSelectedContacts{}} affected parties hosting \empirical{\nPDFsNotified{}} PDFs about their presence.
\end{itemize}

\section{Background and Research Questions} \pagesBudget{1}
\label{sec:background}

\subsection{Background}

In this section, we introduce core concepts, outline the framing of our study and present our research questions.

\subsubsection{\PPDF s}
\label{sec:bg_pPDFs}
\begin{figure}
	\centering
	\includegraphics[width=1\columnwidth]{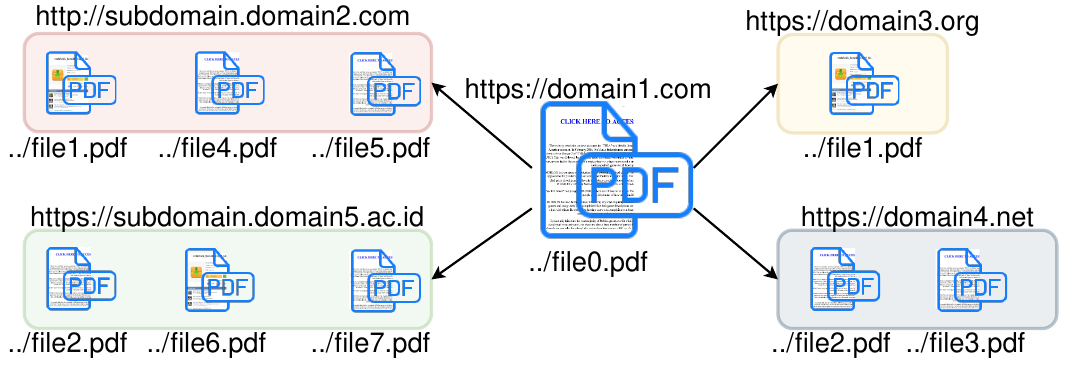}
	\caption{The interconnections between \pPDF s.}
	\label{fig:phishPDFslinks}
\end{figure}

\PPDF s were recently presented in \cite{stivala23from} as fraudulent-looking PDFs functioning as an entry point for a series of Web attacks (phishing, drive-by download, scam, etc), should the victim have clicked on a link embedded in their first page.
In this attack scenario, victim users come across \pPDF s when searching for specific terms on search engines (such as Google and Bing). The PDFs are returned among the first ten search results and seamlessly rendered by the browser upon a click~\cite{stivala23from}.
This high rank in search results is attributed to malicious search engine optimization techniques, which exploit the structure and content of the PDFs to achieve higher rankings. Specifically, \pPDF s display a bait message or image in the first page, which also embeds the link leading to a Web attack, while including 13 to 30 links to different \pPDF s (\cite{stivala23from}) in the following pages to implement backlinking and resources cross-linking to boost SE ranking (\cite{john2011deseo, wu2005identifying}).
Since SEO attacks are the main distribution vector for \pPDF s~\cite{stivala23from}, the rest of this paper focuses on \pPDF s distributed through SEO.
\Cref{fig:phishPDFslinks} displays these interconnections between different \pPDF s.

\subsubsection{SEO Attacks}
\label{sec:bg_seo-attacks}
Search engine optimization (SEO) attacks aim at promoting malicious content high in search results. 
Previous studies established a series of techniques used by attackers in a SEO attack, such as
\emph{keyword stuffing}~\cite{ntoulas2006detecting}, where the promoted content is filled with specific key terms to appear more relevant. \emph{Cross-linking resources}~\cite{wu2005identifying} exploits the link-based ranking algorithms of search engines, where attackers craft a network of ad-hoc resources and cross-link them to influence the ranking of promoted resources. Finally, attackers 
\emph{use benign websites} to host the cross-linked resources~\cite{john2011deseo} as their good reputation positively influences the final ranking.

\subsubsection{Websites Supporting Web Attacks}
Many works discussing URL maliciousness consider a website serving malicious content as either being owned by the attacker or compromised~\cite{de2021compromised, maroofi2020comar, nappa2013driving}.
However, this perspective overlooks scenarios where the attacker neither registers a new domain nor compromises a third party's domain, but rather get assigned a domain from a hosting provider (e.g., a free subdomain). This possibility became feasible with the availability of inexpensive (if not free) services offered by hosting providers, possibly without thorough registration checks. These services may include free object storage, free subdomains for E-commerce websites, or online marketplaces.
When such infrastructure is used for malicious purposes, it is technically incorrect to call it ``compromised'' since the attacker did not exploit the software stack running at that origin. We thus use the broader term \textit{abused infrastructure} to indicate a large amount of websites, potentially managed by a single provider or part of the same hosting service, whose usage is inappropriate, often illicit, resulting in significant harm to the owner and its users.
In the context of \pPDF s, the supporting infrastructure is the ensemble of websites, services and providers whose resources are being misused by attackers to host \pPDF s, such as \texttt{domain1.com}, \texttt{subdomain.domain2.com}, \texttt{domain3.org} and \texttt{domain4.net} in \Cref{fig:phishPDFslinks}.

\subsection{Scope and Contributions}
In this section, we first present our research questions, then, we outline the contributions and framing of this study with respect to a recent work in this field.

\subsubsection{Research Questions}
The overarching goal of this study is to observe the web infrastructure abused for the distribution of \pPDF s, investigating specific properties concerning its volume and evolution in time.
The first challenge we undertake (\textbf{Research Question 1}) is to understand its composition in terms of hosts or services, for example by identifying Autonomous Systems or any specific hosting services involved, and to which extent. We tackle this research question in $\S$ \ref{sec:id_host_prov}. 
Next, we ask ourselves how attackers acquire upload capabilities to these domains (\textbf{Research Question 2}). Specifically, we look for security-related properties, as the presence of outdated, vulnerable or misconfigured software components which might have been exploited by attackers to gain the ability of uploading \pPDF s. We investigate multiple security properties and report our findings in $\S$ \ref{sec:security_indicators}.
Following, we focus on the duration and volume of the abuse (\textbf{Research Question 3}). We define the duration of abuse by monitoring the online status of all \pPDF s in our dataset with the granularity of a single day ($\S$ \ref{sec:duration-of-abuse}), and its volume by observing the distribution of \pPDF s over the types of hosting we previously identified ($\S$ \ref{sec:campaigns_intro}). 
Lastly, we focus on measures that could be taken to help mitigating the spread against \pPDF s, ultimately protecting users and improving the security of the abused hosts. 
Existing protection methods, as blocklists, provide limited protection for users ($\S$ \ref{sec:anti-phishing-entities}), thus, we evaluate the effectiveness of responsibly disclosing the issue to affected parties (\textbf{Research Question 4}) ($\S$ \ref{sec:vuln_notif}), observing as impact indicators both the number of PDFs that were cleaned up and the domains that did (or did not) see any further upload.

\subsubsection{Contributions}
The closest work to ours is a recent study by Stivala et al.~\cite{stivala23from}. In the following, we elaborate on the differences between the two works, outlining our contributions.
Our work expands on the findings in~\cite{stivala23from} by shifting the focus to the \textit{abused infrastructure} hosting \pPDF s. Our investigation centers on identifying hosting types (RQ1), gathering evidence of upload methods (RQ2), and the impact of responsible disclosure (RQ4), differently from \cite{stivala23from} which focuses on PDF characteristics and distribution methods.
To enhance our understanding of real-time abuse monitoring (RQ3), we introduce two datasets: \dsVT{} and \dsFromUrl{}, where \dsVT{} serves as source to build \dsFromUrl{}, and \dsFromUrl{} allows for direct real-time analysis. Our \dsVT{} is newer and three times larger than \cite{stivala23from}’s dataset, with no temporal overlap nor shared samples.
When assessing the volume and duration of this phenomenon, a shared point of investigation, we focus on live hosts rather than on PDFs: the \texttt{.pdf} links in the \dsVT{} enable direct, real-time abuse monitoring--an aspect not studied in ~\cite{stivala23from} and significantly different from observing VT uploads~\cite{stivala23from}. 
Finally, as a technical improvement, we created a new ML model for clustering, reducing latency and human bias.

\section{Dataset and Pipeline} \pagesBudget{0.7}
\label{sec:dataset-tool}

\subsection{Main and Seed Datasets}
Answering our research questions requires knowledge of the hosts serving \pPDF s, for example in the form of a list of URLs leading to these PDFs.
A source of URLs is given by \pPDF s themselves, as \pPDF s include URLs to other \pPDF s as backlinks (see $\S$ \ref{sec:bg_pPDFs}, \ref{sec:bg_seo-attacks}, and \Cref{fig:phishPDFslinks}).
We leverage this property to construct a first dataset of \pPDF s, the \dsVT{}, acting as source of URLs to other \pPDF s.
By visiting these URLs and downloading the corresponding PDFs we build the dataset for this study, \dsFromUrl{}. The inclusion of a downloaded PDF to the \dsFromUrl{} (as well as to the \dsVT{} in the previous step) is subject to the evaluation of SEO-specific properties (detailed in $\S$ \ref{sec:data_quality} below), ensuring that no benign or non-\pPDF{} is included.

\subsubsection{Data Collection}

Our starting dataset, \dsVT{}, counts \empirical{\totUniquePDFs{}} PDFs with unique SHA-256 signatures, covering a period of \empirical{\durationEntireStudy{}} months (from \empirical{March 14th, 2022} to \empirical{June 26th, 2023}).
The nine-month gap between the start of our study and the end of Stivala et al.'s raises questions about whether those \pPDF s are still online and part of an attack campaign, which we address by collecting up-to-date \pPDF s provided by two industrial partners, who retrieve them from VirusTotal.
The two partners contribute unevenly, accounting for \empirical{\percPDFsInquest{}\%} and \empirical{\percPDFsCisco{}\%} of the entire dataset, respectively.

We start downloading PDFs to construct the \dsFromUrl{} after a \empirical{\durationPrelimStudy{}}-month setup phase.
This second data collection lasted \empirical{\durationStudy{}} months, during which we monitored \empirical{\totMonitoredLinksFromUrl{}} \texttt{.pdf} links.
URLs that are unreachable, do not serve PDFs, or serve non-\pPDF s are discarded, resulting in \totPDFsFromUrl{} URLs that returned a \pPDF{} at least once during the Main phase of the study.
\Cref{tab:vt_dataset} reports the number of \pPDF s in the \dsVT{} and the number links extracted from them, as well as the number of \pPDF{} observed online. $\S$ \ref{sec:pdf_analysis_seo_metric} reports the implementation steps behind our data collection.

\begin{table}
	\centering
	\footnotesize
	\begin{tabular}{l | c | r  r }
		\toprule
		& DS & Setup Phase                 & Main Study                  \\
		\midrule
		Start                                                  &   & 2022-03-14                  & 2022-06-22             \\
		End                                                    &   & 2022-06-21                  & 2023-07-26                    \\
		\midrule
		PDFs                                                & $\square$ & \totPDFsVTPrelimStudy{}       & \totPDFsVTStudy{}        \\
		\hspace*{3mm}of which SEO                           & $\square$ & \totSeoPDFsVTPrelimStudy{}    & \totSeoPDFsVTStudy{}     \\
		Extracted \texttt{.pdf} links                       & -         & \totUriPrelimStudyVTSeoUris{} & \totMonitoredLinksFromUrl{}  \\ %
		\hspace*{3mm}of which online \& SEO                 & $\blacksquare$ & -                        & \totPDFsFromUrl{}                \\
		\bottomrule
	\end{tabular}
	\caption{Volume of unique PDFs in \dsVT{} ($\square$) and \dsFromUrl{} ($\blacksquare$), and unique \texttt{.pdf} links extracted from them.}
	\label{tab:vt_dataset}
\end{table}

\subsubsection{Filtering Criteria}
\label{sec:data_quality}

We implement two filtering criteria to limit the inclusion of benign or non-\pPDF s in our datasets, in line with prior works~\cite{stivala23from}.
Identifying \pPDF{} involves verifying the presence of SEO characteristics, which are not visible from the \texttt{.pdf} URLs but can be observed by inspecting the PDF structure and content (see $\S$ \ref{sec:bg_seo-attacks}).
We thus download and parse PDFs, ensuring the presence of SEO characteristics in two ways before adding them to the \dsVT{} and \dsFromUrl{}.
These criteria (hereinafter \textit{SEO metric}) ensure the presence of at least five \texttt{.pdf} links in total, relaxed from the original ten, and a mean number of at least one \texttt{.pdf} link per page, consistent with~\cite{stivala23from}. This change was due to our different data sources (VirusTotal and backlinks in \pPDF s) where the distribution of benign documents is much lower than that of search engines like Google and Bing.
The lower threshold is designed to include a large number of  \pPDF{} documents while minimizing false positives.
$\S$ \ref{sec:eval_seo_metric} reports on the accuracy of this metric.

\subsection{The \toolName{} Pipeline}
\label{sec:pipeline}
\begin{figure}
	\centering
	\includegraphics[width=1\linewidth]{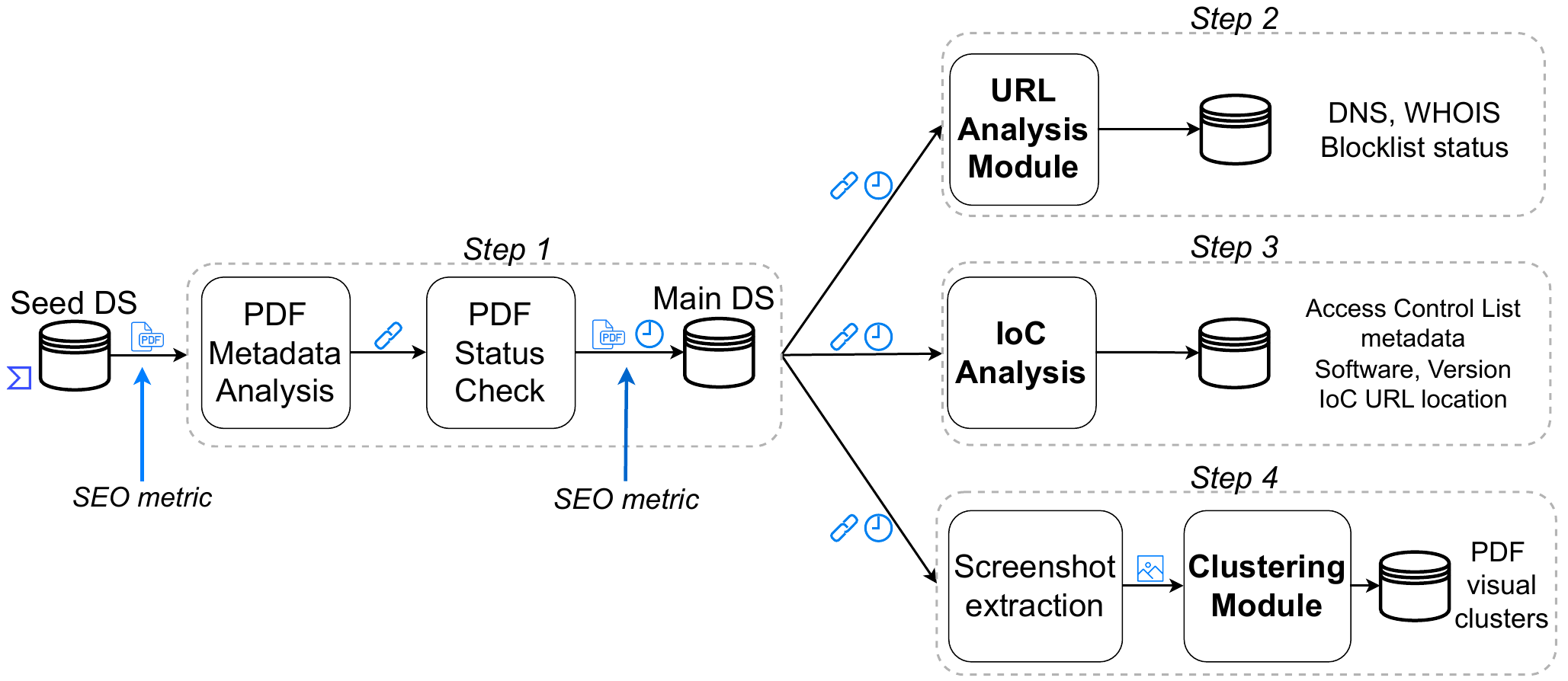}
	\caption{\toolName{} modules and I/O data connections.}
	\label{fig:pipeline}
\end{figure}

The initial \empirical{three-month} setup phase are necessary to build \toolName{}, shown in \Cref{fig:pipeline}, a modular pipeline running daily in real-time. \toolName{} ingests and processes millions of tiny PDF-related pieces of information from various sources every day. When ``mashed'' together, these pieces reveal valuable insights into the \pPDF{} threat. %
We release the code of \toolName{} at \href{https://github.com/emerald1010/hosts-supporting-clickbait-PDFs}{https://github.com/emerald1010/hosts-supporting-clickbait-PDFs}.

The first module (\textit{Step 1}) processes the PDF binaries received from our industry partners, extracting useful metadata such as the embedded URLs. These are fed into the \pdfCheckPipeline{}, which visits them and defines their online or offline status. These pairs \texttt{(URL, datetime\_information)} constitute the basis of the \dsFromUrl{} and are the input of all following modules. Specifically, we fetch DNS and WHOIS of all URLs in the \dsFromUrl{} (\textit{Step 2}), visit the websites hosting online PDFs looking for indicators of compromise (\textit{Step 3}) and, finally, download online PDFs and extract the screenshot of the first page (\textit{Step 4}) to determine the groups of visual baits.
The modules are orchestrated and monitored via an instance of Apache Airflow.
Following, we detail the behavior of each module.

\subsubsection{PDF Analysis Module}
\label{sec:pdf_analysis_seo_metric}
We begin by choosing PDFs from the \dsVT{} that meet the SEO metric. Next, we extract their URLs and metadata (\textit{PDF Metadata Analysis}), and subsequently verify the online status of those URLs leading to a PDF (\pdfCheckPipeline{}). 
In the \textit{PDF Metadata Analysis}, the URLs are obtained by reconstructing the PDF tree with a modified version of the open-source library \texttt{peepdf}~\cite{peepdf}, by navigating the tree breadth-first looking for nodes encoding URLs (e.g., \texttt{URI}) or whose parent node's attributes include \texttt{Subtype Link}, \texttt{Rect} and either \texttt{Type Annot} or \texttt{Type A}.
This approach was preferred to a simple string matching (e.g., looking for \texttt{http://}-like strings) as it allows extracting URLs in compressed streams.
Lastly, we collect the document title by inspecting its \texttt{Document Information Dictionary} and obtain the screenshot of the first page via the \texttt{Poppler}~\cite{poppler} utility using 150 dots per inch.

\pdfCheckPipeline{} consists of a module performing daily HTTP requests to the extracted \texttt{.pdf} links, de-facto recording the uptime of each linked PDF. 
We monitor each link on a daily basis starting from the day of its initial observation, and continue until it remains offline for three consecutive days.
A URL is considered offline when its \texttt{Content-Type} header is different from \texttt{application/pdf}, or if it returns a status code $>=300$. 
To reduce the load on the target domains, we initially perform \texttt{HEAD} requests, and proceed with a \texttt{GET} only if the above criteria are met. Moreover, we store the linked \pPDF{} on the first visit.
We also included the use of numerous VPN endpoints to check that a given domain is not blocklisting us before marking its URLs as offline. 
\pdfCheckPipeline{} became operative on \empirical{June 22nd, 2022}, marking the start of the Main phase of our study (no PDF was downloaded prior to this date).
$\S$ \ref{sec:limitations} discusses possible limitations of this approach and $\S$ \ref{sec:ethics} discusses the measures we took to reduce the load of our analyses on target websites.
Before adding new PDFs into the \dsFromUrl{}, we ensure they meet the \textit{SEO metric}, and then we reapply the \textit{PDF Metadata Analysis}.

\subsubsection{URL Analysis Module}
\label{sec:fromurl_module}
In this step we perform analyses on the extracted URLs. 
We collect DNS records of each fully-qualified domain name (FQDN) actively serving \pPDF s, extract its IP and fetch the corresponding WHOIS record, including Autonomous System numbers.
Next, we collect the blocklist status of each extracted \texttt{.pdf} link, using Google SafeBrowsing (pre-installed on more than 84\% of users' browsers~\cite{statista_browser_market_share}) and VirusTotal, popular both in research and in industry (see, e.g.,~\cite{oest2019phishfarm, zhu2020measuring}) as reference.

\subsubsection{Indicators of Compromise Collection Module}
The collection of indicators of compromise is a multi-faceted procedure which comprises different analyses depending on the target host. It is performed by two sub-modules collecting evidence of vulnerable or misconfigured software components.

The first module collects indicators linked to the presence of software components and plugins running on the server-side by visiting with a full-fledged Chrome browser the homepage of a domain actively serving \pPDF s. When loading the page, the browser waits up to 15 seconds, intercepting all network requests happening in the background. This functionality is similar to that realized by~\cite{wptagent}, which we incorporate for easier interaction with the Linux Traffic Interface. 
We then process the network traces applying a rule-based approach (we integrate that of~\cite{wappalyzer} for simplicity, similarly to~\cite{de2021compromised})  to obtain information on the web server (e.g. Apache), programming languages (e.g., PHP), hosting panels (e.g., Plesk), web application framework (e.g., Wordpress) and add-ons (as WordPress Themes and Plugins).

Our second module is a custom vulnerability scanner developed to verify the presence of misconfigured or vulnerable components which may lead to file upload. The scanner visits pre-selected URL paths which we observe are indicators signalling the presence of a component allowing file upload. $\S$ \ref{sec:cksource_vulns} details the inner workings of this component.
In case no evidence could be collected we trigger additional analyses for this FQDN, where the Chrome browser visits $n\leq20$ random pages extracted from the homepage of the domain to possibly observe additional software components.

\subsubsection{Clustering Module}
\label{sec:ML_detector}
\PPDF s can be clustered with respect to the visual deceit (e.g., position and aspect of their bait elements) shown on the first page~\cite{stivala23from}.
Previous work identified \empirical{44} clusters using a Deep Learning approach based on Convolutional Neural Networks (CNNs).

We develop our own CNN model to perform feature extraction, creating a feature space where visually-similar samples are mapped close to each other.
The model takes a screenshot of the first page of each document and returns a 32-dimensional vector denoting its position in the new feature space.
We create a training set starting from the one provided by~\cite{stivala23from}. We performed data cleaning when necessary, removing outliers and filtering or remapping elements to new groups based on their similarity. 
Finally, we augment it with more recent data from our data feeds, obtaining a total of \empirical{\samplesInTraining{}} training samples divided into \empirical{\campaignsInTraining{}} groups.
Next, we use a semi-hard triplet selection process and the triplet-loss function to train the model weights (see~\cite{schroff2015facenet}).
With this model, we extract a feature vector for each PDF and then apply DBSCAN~\cite{ester1996density} for clustering.
To reduce manual intervention, we incorporate pre-labeled samples, or ``anchors'', into the pool of unseen documents. 
This way, we can automatically label the clusters based on the group of anchors they contain. If multiple anchors are associated with the same computed group, we re-cluster its samples using a smaller $\epsilon$ with DBSCAN until the conflict is resolved. Human intervention is only required when our \textit{Clustering module} identifies a new cluster.
\Cref{sec:appendix-clustering} provides further details on the model and clustering procedure.

\section{Characterizing Support Infrastructure} 
The goal of this section is twofold. 
Firstly, we examine the host and service composition, seeking similarities among hosts. 
Addressing this early on in the setup phase enables us to conduct specific analyses later on for these host types, which we study during the main phase.
To tackle \textbf{RQ 1}, we investigate the network properties (Autonomous System, DNS lookup, URL) of the \empirical{\totUriPrelimStudyVTSeoUris{}} URLs extracted from \dsVT{} (backlinks leading to to \pPDF s, see $\S$ \ref{sec:bg_pPDFs}). Our analysis of these properties ($\S$ \ref{sec:id_host_prov})
reveals the presence of large groups of hosts with similar traits. Specifically, we observe \empirical{three} different types of hosting, covering \empirical{\totHostingProviders{}} eTLD+1s.
Next, in $\S$ \ref{sec:security_indicators} we run ad-hoc analyses on the websites serving \totPDFsFromUrl{} live \pPDF s during the Main phase of the study.
We identify \empirical{six} plugins and \empirical{two} web frameworks facilitating file upload, and \empirical{\domsOutdatedComponents{}} origins hosting outdated software components, answering \textbf{RQ 2}.

\subsection{Analysis of Network Properties} \pagesBudget{1}
The goal of this section is to identify whether certain hosts within the supporting infrastructure share similar features, which we define in terms of network properties.

To find out if and which components make up the supporting infrastructure, we conduct an exploratory analysis of the \dsVT{} backlinks.
Since attackers target large amounts of websites having the same security flaw (see, e.g., ~\cite{vasek2015hacking, zhang2012poisonamplifier, moore2009evil}) we analyze our data to find large groups of hosts sharing similar network properties.
Our approach does not aim at identifying hosting provider \textit{organizations}~\cite{soussi2020feasibility} but groups of similar Web hosts targeted by attackers. 

\label{sec:id_host_prov}
\subsubsection{Methodology}
\begin{table}
	\centering
	\footnotesize
	\begin{tabular}{l  r V{3} l  r }
		\toprule
		Autonomous System & \# FQDNs & Autonomous System & \# PDFs \\
		\midrule
		WEEBLY, US        &  \numprint{41483}	& 		WEEBLY, US	            & \numprint{241851}\\ 
		\textit{AMAZON-02, US}	&  \numprint{9222}	& 	\textit{AMAZON-02, US}	& \numprint{142200}\\ 
		WILDCARD-AS	      &  \numprint{5351}	& 		CDN77 \textasciicircum\_\textasciicircum, GB	& \numprint{59213}\\ 
		\textit{GOOGLE-2, US}	&  \numprint{4301}	&   \textit{CLOUDFLARENET}	& \numprint{57156}\\ 
		ZETTA-AS, BG	  &  \numprint{4091}	& 		\textit{GOOGLE-2, US}	& \numprint{46264}\\ 
		AUTOMATTIC, US	  &  \numprint{1556}	& 		UNIFIEDLAYER	        & \numprint{37504}\\ 
		\textit{CLOUDFLARENET}	&  \numprint{1363}	& 	OVH, FR	                & \numprint{34974}\\ 
		OVH, FR	          &  \numprint{1141}	& 		\textit{GO-DADDY-CO}	& \numprint{31080}\\ 
		IWEB-AS, CA	      &  \numprint{1097}	& 		\textit{ARUBA-ASN, IT}	& \numprint{25731}\\ 
		UNIFIEDLAYER	  &  \numprint{1086}	& 		FASTLY, US	            & \numprint{25703}\\ 
		\bottomrule 
	\end{tabular}
	\caption{Top ten Autonomous Systems sorted by number of FQDNs (on the left) and by the number of PDFs (on the right). The two lists report different AS names depending on their rank determined by the sorting criterion.}
	\label{tab:as_data_uri_netloc}
\end{table}

\begin{figure*}[h]
	\centering
	\begin{subfigure}{0.46\linewidth}
		\centering
		\includegraphics[width=0.95\linewidth]{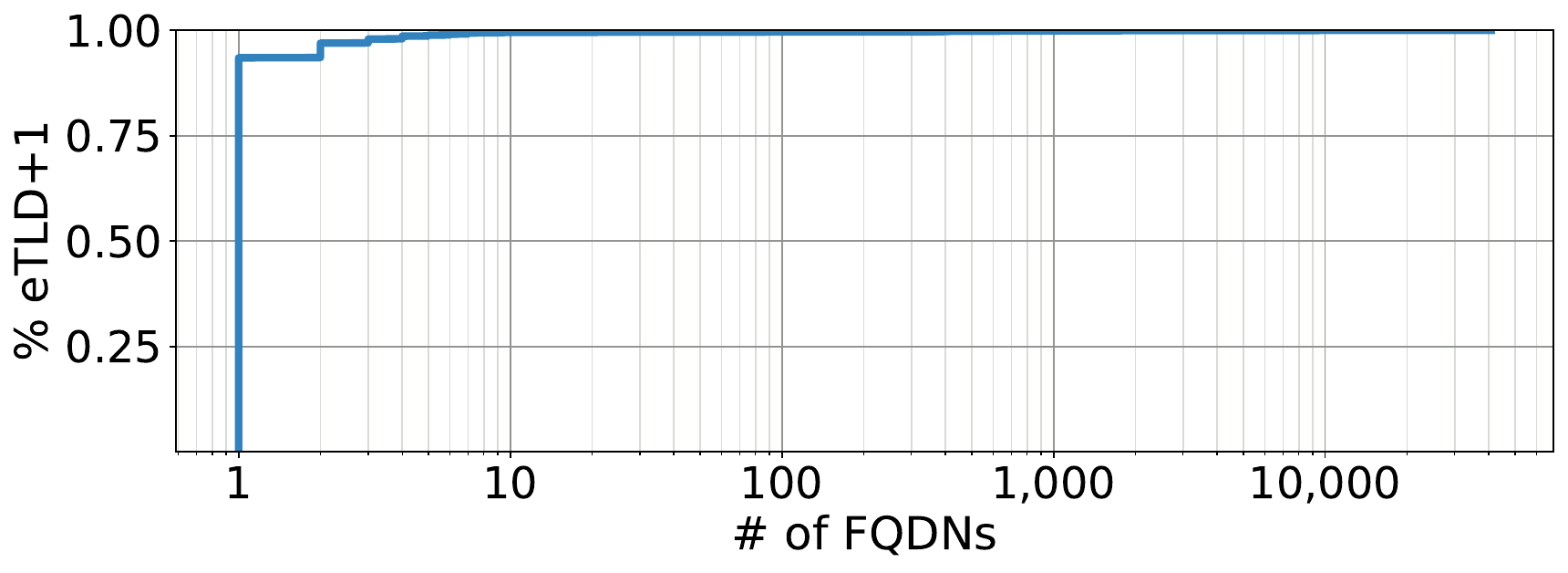}
		\caption{}
		\label{fig:cdf_activity_subdomains}
	\end{subfigure}
	~~
	\begin{subfigure}{0.45\linewidth}
		\centering
		\includegraphics[width=0.95\linewidth]{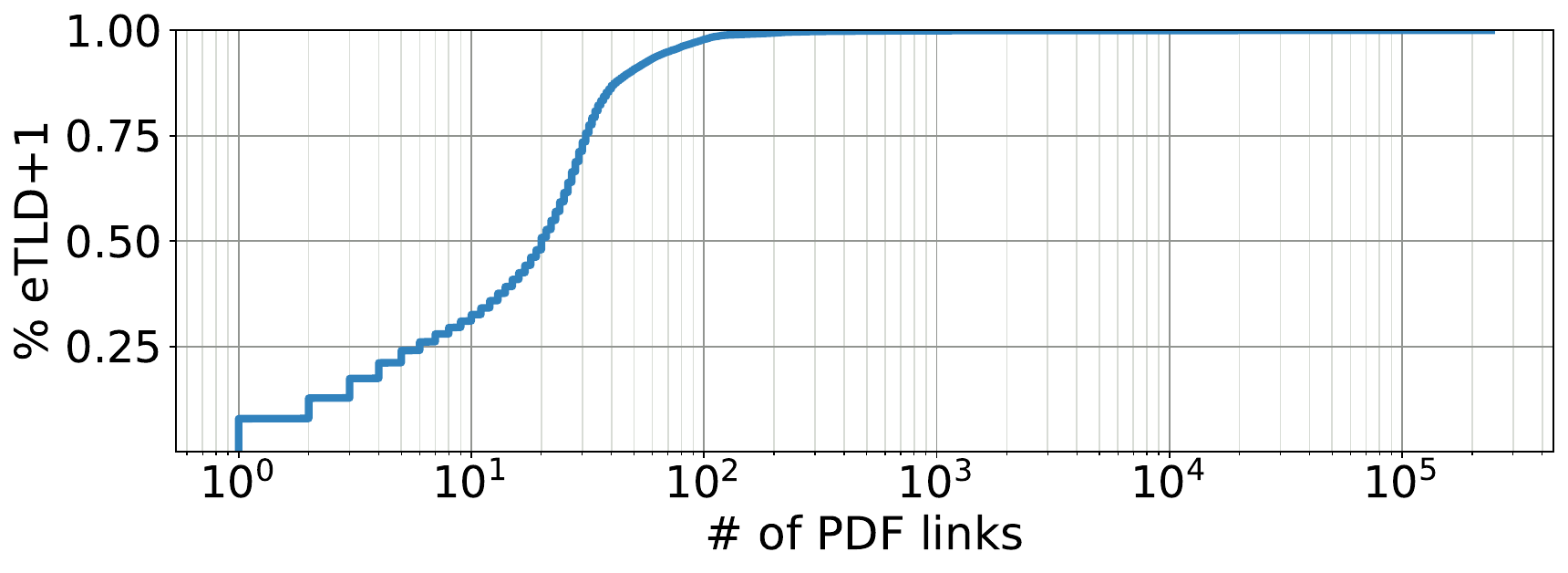}
		\caption{}
		\label{fig:cdf_activity_links}
	\end{subfigure}
	\caption[]{(a) Distribution of FQDN per eTLD+1. (b) Distribution of \texttt{.pdf} links per eTLD+1. Data from the \textit{Setup phase}.}
\end{figure*}

We focus on those indicators that can either be observed directly (e.g., domain name) or obtained via well-established channels (e.g., DNS queries).
For example, given a URL \texttt{http://babemozigu.weebly.com/dir/file.pdf} we extract its FQDN (\texttt{babemozigu.weebly.com}) and its eTLD+1 (\texttt{weebly.com}), or ``domain root''. 
We obtain the IP address and the Autonomous System (AS) for each FQDN from the respective DNS and WHOIS records.
For readability purposes, we aggregate different AS names belonging to the same company (e.g., \texttt{CLOUDFLARENET, US} and \texttt{CLOUDFLARESPECTRUM Cloudflare, Inc., US}, in italics) and report in \Cref{tab:as_data_uri_netloc} two distinct lists of the ten most affected ASes, independently sorted by number of unique FQDNs and by number of observed \pPDF s. 

We noticed a significant difference in the order of ASes between the two lists. For instance, the ASes for \texttt{Weebly}, \texttt{Wilcard}, and \texttt{Zetta-AS} (first, third, and fifth ASes) were found to be the most frequently abused in terms of FQDN, but their overall rank differs considerably when sorting them by number of \pPDF s.
\Cref{fig:cdf_activity_subdomains} shows the distribution of FQDNs per domain root. The graph shows a sharp increase, indicating that the majority of domain roots (\empirical{96\%}) have either no subdomain or just one subdomain. However, a small percentage ($< 0.01\%$) of domain roots have \empirical{ten} or more subdomains.
To further analyze this, we set an empirical threshold of \empirical{100} FQDNs per domain root and manually investigate the resulting \empirical{\amtWebhSameProvs{}} domain roots. These eTLD+1s represent \empirical{97\%} of the domain roots with at least one subdomain. For example, \texttt{babemozigu.weebly.com}, \texttt{babewepuk.weebly.com}, and \texttt{babexunerasosib.weebly.com} are among them.

In the right column of \Cref{tab:as_data_uri_netloc}, we present different ASes based on the number of served \pPDF s. The distribution of PDF files across domain roots (\Cref{fig:cdf_activity_links}) shows that most eTLD+1s host a maximum of \empirical{100} \pPDF s, while only \empirical{2\%} of the domains serve more than that. We adopted a conservative approach to identify candidate domain roots by using the number of \texttt{.pdf} links as a criterion. As uploading a large amount of PDFs to a compromised website is easier than obtaining free subdomains, we set an empirical threshold of \empirical{\numprint{5000}} PDF links per eTLD+1. We manually investigated the domain roots that exceeded this threshold in terms of PDF volume. 

\subsubsection{Results}
\label{sec:hosting_services}
\begin{figure}
	\centering
	\includegraphics[width=0.60\linewidth]{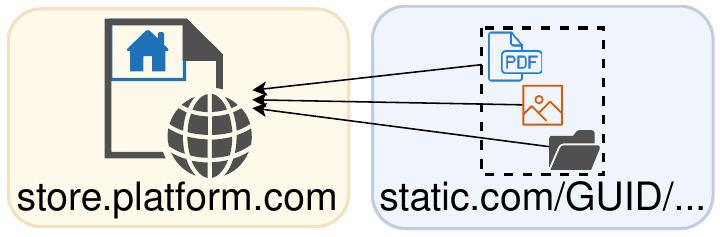}
	\caption{Example showing static resources residing on a different domain (PDFs in the \webhDiff{} category).}
	\label{fig:sketch-webhdiff}
\end{figure}
This procedure identified \empirical{26} unique eTLD+1s. 
We confirm the existence of specific hosting services running on that domains by conducting a separate market research for services exhibiting similar characteristics. As a result, we identified \empirical{three} services running on these eTLD+1s, namely \objst{}, \webhDiff{} and \webhSame{}, which we explain below.

\objst{} is a hosting service that manages unstructured data, such as PDFs, as individual units, or \textit{objects}, stored in a single location~\cite{objectStorage_Google, objectStorage_Cloudflare}. The URLs of these objects include strings resembling unique identifiers, either as subdomains or in the URL path. Although these origins cannot be browsed, files can be retrieved using known URLs.
We found \empirical{one} domain root belonging to this category, whose service includes a free tier accessible after thorough checks (e.g., providing a valid credit card number).

\webhDiff{} origins exhibit a filesystem structure that resembles that of \objst{} services, where PDFs (and other static resources) reside on a separate origin from where the main website operates, as depicted in \Cref{fig:sketch-webhdiff}. 
Through our manual analysis, we were able to link all but two of them (\texttt{sqhk.co} and \texttt{f-static.net}) to a specific hosting service, such as E-commerce marketplaces or Shared hosting.
During our market research on online hosting services, we discovered one entity using multiple eTLD+1s, as \texttt{s123-cdn-static-a.com} and \texttt{s123-cdn-static-b.com}. In our dataset, we identified \empirical{\nManuallyAdded{}} such instances and included them in this category.

In \webhSame{} services, multiple websites run on the same server. The services running on those domains observed in our data offer affordable options, including free subdomains or automated website building, an online service that enables users to create websites without coding skills by combining pre-designed modules.
We verified that these services allow users to publish a website without requiring a credit card or a valid email address. 

We perform two extra checks to ensure that no other hosting service with a lower volume of abuse went undetected. First, we investigate the remaining FQDNs via a third-party web analytics service~\cite{similarweb}, observing \empirical{\nSimilarWeb{}} additional domain roots classified as \textit{\catFour{}}. We verify the correctness of this label before adding them to our \webhSame{} group.
Additionally, we checked our URLs against a manually curated list of hosting services, finding \empirical{two} URL matches for \texttt{digitaloceanspaces.com} (DigitalOcean) and \empirical{four} URL matches for \texttt{storage.googleapis.com} (Google). However, we do not include them in our further analyses as the volume of URLs for these two providers is negligible with respect to that of other \objst{} providers identified by our methodology (e.g. Amazon, \empirical{\nUrlsAmzSetupPhase{}}).
\Cref{tab:urls_per_ht} reports the volume of \pPDF s per hosting type and category, while exhaustive details on the identified hosting services (as eTLD+1, volume of \pPDF s and FQDNs) are reported in \Cref{tab:IDed_providers} (Appendix).
We observe that the coverage of our websites provided by~\cite{similarweb} is limited (\empirical{\percTldsWCategoryData{}\%} of all domain roots), which might be explained by the low rank of some websites or by their offline status.
In the remaining, we refer to websites in none of the groups \objst{}, \webhDiff{}, or \webhSame{} as \uncat{}.

\begin{table}
	\centering
	\footnotesize
	\begin{tabular}{l | r}
		\toprule
		Hosting Type & \# URLs                            \\
		\midrule
		\objst{}     & \empirical{\allUniqueURLsObjst{}}     \\
		\webhDiff{}  & \empirical{\allUniqueURLsWebhDiff{}} \\
		\webhSame{}  & \empirical{\allUniqueURLsWebhSame{}} \\
		\midrule
		Remaining URLs (\uncat{}) & \empirical{\allUniqueURLsUncat{}}    \\
		\hspace{3mm}\catOne{}     & \empirical{\catOneVolume{}}    \\
		\hspace{3mm}\catTwo{}     & \empirical{\catTwoVolume{}}    \\
		\hspace{3mm}\catThree{}     & \empirical{\catThreeVolume{}}    \\
		\hspace{3mm}\catFour{}     & \empirical{\catFourVolume{}}    \\		
		\hspace{3mm}\catFive{}     & \empirical{\catFiveVolume{}}    \\
		\hspace{3mm}Remaining categories     & \empirical{\othCatVolume{}}    \\
		\hspace{3mm}No category found by~\cite{similarweb}     & \empirical{\noCatVolume{}}    \\
		\bottomrule 
	\end{tabular}
	\caption{Number of URLs to \pPDF s over hosting types or website categories. Data from the \textit{Main phase}.}
	\label{tab:urls_per_ht}
\end{table}

\subsubsection{Takeaways}
In this section we addressed \textbf{RQ 1} by scrutinizing observable properties of URLs hosting \pPDF s.
Our methodology identified a total of \empirical{\totHostingProviders{}} domain roots (\empirical{26} via analysis of network properties, \empirical{\nManuallyAdded{}} via manual analysis, and \empirical{\nSimilarWeb{}} via a third-party service~\cite{similarweb}), which we have verified correspond to existing hosting types and services. 
For the scope of this paper, we organized them in \empirical{three} broad groups, \objst{}, \webhDiff{}, and \webhSame{}. Note that these names might not cover all the extensive services provided by major providers. For instance, \webhSame{} might involve Website Builder services, along with managed and unmanaged shared hosting.

\subsection{Indicators of Compromise}
\label{sec:security_indicators}
In this section, we investigate factors which may have facilitated the upload of \pPDF s on the abused hosts, answering \textbf{RQ 2}.
Our analyses are tailored to the characteristics of each hosting type, investigating Access Control Lists, presence and up-to-date status of software components related to website abuse, and plugins which we observed to lead to file upload. %
We observe the strong presence of outdated and vulnerable components on \uncat{} websites, while \webhSame{} domains present a bare software stack which is rarely outdated.
Finally, we summarize our main findings.

\subsubsection{Experimental Setup}
Different hosting types expose distinct properties, requiring the development of custom analyses modules for each type.

Firstly, when their URL is requested (e.g., via \texttt{HTTP GET}), \objst{} hosts return ``data units'', and authorized users can upload new data via protocols specified by the service provider.

Next, we consider \webhDiff{} providers and observe that domains in this category return an HTTP status code \texttt{403} when requesting the base path (``\texttt{/}'') or any path segment preceding a PDF file. In fact, their filesystem structure cannot be inspected via simple \texttt{HTTP} requests, similarly to \objst{} origins.
Collecting data on the respective ``storefront'' of \webhDiff{} origins is impossible because systematically linking \webhDiff{} origins to their respective homepage domains is infeasible (see \Cref{fig:sketch-webhdiff}). Consequently, we removed all domains belonging to this category from the analysis.

Conversely, websites belonging to the \webhSame{} or \uncat{} categories can be inspected via regular crawling.
We determine the presence of outdated or vulnerable components in two ways. First, we compile a list of server-side software components that previous works found to be connected to Internet abuse. These are: \textit{(i) }type of web application, specifically CMSes and E-commerce software; \textit{(ii)} their version (see, e.g.,~\cite{goethem2014large, tajalizadehkhoob2017herding}); \textit{(iii)} a list of plugins and themes, as the ones for WordPress, when applicable (see, e.g.~\cite{kasturi2022mistrust, vasek2015hacking}). \textit{(iv)} the presence of Unrestricted File Upload vulnerabilities, as highlighted to be used in conjunction with SEO attacks~\cite{john2011deseo}.
Second, we performed a manual analysis of selected URLs, which led to the identification of \empirical{\identifiedVulnerableComponents{}} additional components linked to file upload, for which we develop a custom scanner.  

We follow best practices and disclosure guidelines in these analyses. Due to ethical concerns, we develop non-intrusive analyses looking for indicators of compromise (hereinafter IoCs), refraining from sending \texttt{POST} requests to verify vulnerabilities when this would trigger a state change on the target website.

\subsubsection{Misconfigured S3 Buckets}
\label{sec:misconfigureed_buckets}
The only \objst{} service in our dataset corresponds to Amazon's Simple Storage Service. Thus, our analysis of \objst{} websites is based on the collection of metadata on S3 buckets permissions.
We develop our S3 scanner module relying on a popular library~\cite{s3scanner} on top of the AWS SDK. Similarly to~\cite{continella2018there}, we proceed with the inspection of each bucket, collecting Access Control Lists (ACLs) and bucket contents when possible. For ethical reasons, we do not try to write any file to the buckets.
We observed that a bucket may still exist even if one or more referenced PDFs are not online, thus, we feed the S3 scanner module all \objst{} links, regardless of their online status.
We probed \empirical{\awsBucketsScanned{}} unique buckets in total, obtained from \empirical{\urisScannedTot{}} links, where \empirical{\awsExistingBuckets{}} were reachable at the time of scanning, while the remaining ones raised an error (e.g., \texttt{NoSuchBucket} or permission denied). 
We find that \empirical{\awsBucketsACL{}} of them have a readable Access Control List, where \empirical{\percAwsBucketsFC{}\%} of the buckets leave \texttt{Full Control} permissions, \empirical{\percAwsBucketsW{}\%} of the buckets leave \texttt{Write} permissions, and \empirical{\percAwsBucketsReadACP{}\%} of the buckets allow to read a bucket's ACL (\texttt{READ\_ACP} permission) to unauthenticated users.

\subsubsection{Outdated Software Components}
\label{sec:outdated_sw}
\begin{table}
	\centering
	\footnotesize
	\begin{tabular}{ll r r}
		\toprule
		SW Category &                  SW Name & \# versions & \# FQDNs \\
		\midrule
		CMS &             WordPress &     188 &   4,041 \\
		CMS &                Joomla &       3 &    209 \\
		CMS &                Drupal &       3 &    112 \\
		Ecommerce &           WooCommerce &     150 &   1,310 \\
		Ecommerce &  EasyDigitalDownloads &      11 &     24 \\
		Ecommerce &               Magento &       1 &      4 \\
		Prog. language &                   PHP &     280 &   8,206 \\
		Web servers &                Apache &      40 &   1,884 \\
		Web servers &                 Nginx &      68 &   192 \\
		Web servers &                   IIS &       7 &    438 \\
		WP plugins &             Yoast SEO &     193 &   1,463 \\
		WP plugins &           WooCommerce &     150 &   1,310 \\
		WP plugins &             Revslider &     115 &    623 \\
		WP themes &                 Astra &      56 &    170 \\
		WP themes &       Hello Elementor &       8 &     71 \\
		WP themes &               OceanWP &      29 &     66 \\
		\bottomrule
	\end{tabular}
	\caption{Three most popular outdated software components per category.}
	\label{tab:outdated_sw_components}
\end{table}

Next, we consider \webhSame{} and \uncat{} websites.
We proceed with a two-way approach: first, we collect data on the software components running at all \webhSame{} and \uncat{} websites actively serving PDFs.
When no data point has been collected for a domain, we randomly select $n\leq20$ additional links from its home page and visit them, to increase the probability of triggering and detecting a vulnerable component.

We focus on software components of the following categories: Content Management Systems (CMSs), Ecommerce software, Hosting panels, Web servers, plugins and themes (as those of WordPress), and software components using the PHP programming language.
We visited all FQDNs that served at least one \pPDF{}, i.e., \empirical{\nDomainsCfInStudy{}} websites, and observed indicators relative to the above categories for \empirical{\percCompFrwOverTot{}\%} of them, identifying a total of \empirical{\nSwObserved{}} software components.
Next, we determine outdated software components by comparing their observed version on a target domain to their latest version at the time. 
We observe that most of the domains where this information is available are \uncat{} domains (\empirical{\percDomainsObservationsUnkn{}\%} of the total observations), where more than half of these websites run outdated components. %
Conversely, only \empirical{\percOutdatedWebhSame{}\%} of the software components observed on \webhSame{} domains are outdated. \Cref{tab:outdated_sw_components} reports the most popular outdated components per category.

As a last step, we inspected the network traces of our scanners to determine why no information was collected for a large amount of FQDNs. This inspection revealed that \empirical{\percNoIndicatorDomainsWeebly{}\%} of the websites that did not return any information are \texttt{weebly.com} subdomains, where the crawling was unsuccessful for \texttt{Timeout} errors as the IP was blocked. All the other domain roots were regularly visited by our scanner\footnote{We strived to reduce the load on target websites performing analyses only once per FQDN.}.

\subsubsection{Vulnerable Software Components}
We construct Common Platform Enumeration identifiers~\cite{CPE} using the retrieved software and version information (\empirical{\componentsWVersion{}} software components with version), and query the National Vulnerabiliy Database (NVD)~\cite{NVD} to obtain corresponding CVE information. We enrich this data with vulnerability information from the WPScan Wordpress Vulnerability Database~\cite{wpscan}. 

Among these, we identified \empirical{\vulnComponents{}} software components whose version, at the time of our inspection, was vulnerable.
We filtered out vulnerabilities less likely to be linked with \pPDF s (e.g., buffer overflow) and focused on ``Unrestricted File Upload'' vulnerabilities. In total, we observed \empirical{\nVulnsUfu{}} vulnerabilities of this type affecting \empirical{\ufuComponents{}} software components among those we inspected. 
Among those domains with software and version information, \empirical{\fqdnsAnalyzedForVulns{}} ran a component listed in either the NVD or the WP vulnerability database, and \empirical{\ufuVulnerableDoms{}} of them had a UFU vulnerability, all of them belonging to the \uncat{} group.

\subsubsection{Software Facilitating File Upload}
\label{sec:cksource_vulns}
\begin{table}
	\centering
	\footnotesize
	\begin{tabular}{l  r | r r}
		\toprule
		SW Component    & \multicolumn{3}{c}{\# FQDNs}\\
		& Path IoC & Scanned   & \% vulnerable \\
		\midrule
		KCFinder    & \empirical{\kcfinderPathGcp{}}  & \empirical{\kcfinderCpScan{}}  & \empirical{\vulnKcfinderCpScan{}} \\
		CKfinder    & \empirical{\ckfinderPathGcp{}}  & \empirical{\ckfinderCpScan{}}  & \empirical{\vulnCkfinderCpScan{}} \\
		FCKEditor   & \empirical{\fckeditorPathGcp{}} & \empirical{\fckeditorCpScan{}} & \empirical{\vulnFckeditorCpScan{}} \\
		CKEditor    & \empirical{\ckeditorPathGcp{}}  & \empirical{\ckeditorCpScan{}}  & \empirical{\vulnCkeditorCpScan{}} \\
		\midrule
		Webform     & \empirical{\webformPathGcp{}}   & -  &-\\
		Formcraft   & \empirical{\formcraftPathGcp{}} & -  & -   \\
		\midrule
		SLiMS       & \empirical{\SLiMSeLearnPathGcp{}} & \empirical{\SLiMSeLearnCpScan{}}  & \empirical{\vulnSLiMSCpScan{}}\\
		E-Learning Madrasah & \empirical{\SLiMSeLearnCpScan{}} & \empirical{\SLiMSeLearnCpScan{}} & \empirical{\vulnELearnCpScan{}}\\
		\bottomrule
	\end{tabular}
	\caption{Number of FQDNs running software facilitating file upload, with IoCs found in the URL path or via crawling.}
	\label{tab:domains_per_cp}
\end{table}
An exploratory manual analysis of \webhSame{} and \uncat{} websites revealed the massive presence of specific vulnerable or misconfigured plugins which could be abused to upload files.
In particular, we analyzed the URLs looking for recurring URL path elements on a large scale, with a volume sufficiently large for them to be considered as a deliberate target.
Our intuition comes from the observation that large numbers of URLs can be grouped together by path segments, e.g., \empirical{\uriPathPattern{}} URLs residing on \empirical{\netlocPathPattern{}} different domains share the path segment \texttt{wp-content/plugins/formcraft/}.
A manual analysis of the most common URL path groups (we could confirm \empirical{19} unique URL path patterns inspecting \empirical{194} websites) led to the identification of \empirical{\nGuessedComponents{}} CMS add-ons and \empirical{two} Web frameworks\footnote{The plugins CKEditor~\cite{CkEditorOfficial}, CKFinder~\cite{CKFinderOfficial}, FCKEditor~\cite{FCKEditorSource}, KCFinder~\cite{KCFinderOfficial}, Formcraft~\cite{FormCraftOfficial}, Webform~\cite{WebformOfficial}, and the Web frameworks E-Learning Madrasah~\cite{sutiah2021software, madrasah_article} (shipped with CKFinder) and Senayan Library Management System~\cite{SLiMS}.}, all having associated CVEs or a public exploit in popular repositories (\Cref{app:details-exploits-cves} reports details and vulnerabilities for each component, while \Cref{app:details-path-indicators} lists path segment indicators).

The presence of IoCs in the path of a URL may be an early indicator of the presence of vulnerable software, which however does not exclude the presence of the same vulnerable components on websites whose URL paths do not have such indicators.
We determine that a website runs a vulnerable component by matching the source code and version string of the component against a regular expresssion\footnote{For example, \texttt{FCKeditorAPI=\{ Version:'2.3.2', VersionBuild: '1082'\}}}. We found specific \texttt{.txt}, \texttt{.js}, or \texttt{.html} files exposing plugin versions through exploit repositories, manual inspection of compromised websites, or by inspecting the source code of the \empirical{\identifiedVulnerableComponents{}} components. We compiled a list \empirical{107} possible locations for these files, which our crawler visits.
We ran this analysis for \empirical{four} plugins, i.e., CKFinder, KCFinder, CKEditor and FCKEditor (verifying the vulnerability for the other two plugins was not allowed, as it required sending \texttt{POST} requests.) %
Visiting all \empirical{107} potential IoC locations for the unseen \webhSame{} and \uncat{} websites daily is an expensive operation, not to mention the traffic load imposed on the target websites.
To reduce the dimension of the data in our daily analyses we \textit{(i)} group domains by URL path (i.e., all path segments excluding the file name), as an identical server-side directory structure is a clear indicator of the presence of a shared server-side component, and \textit{(ii)} visit ten randomly-sampled websites per path group.
After two weeks, we inspect the results and remove all potential IoC locations that did not produce any match, lowering their number to \empirical{\cpEndpoints{}}.

We observed \empirical{\domainsObservationsCpScan{}} websites mounting one or more of the four ``CK'' plugins, \empirical{\percVulnCpScanner{}\%} of which were vulnerable. 
It is remarkable that these domains, all marked \uncat{}, actively served a total of \empirical{\numprint{190258}} PDFs. %
We adopted a similar approach to verify the presence of vulnerable components in the SLiMS and E-learning websites. 
\Cref{tab:domains_per_cp} shows the amount of domains whose URL path contains an IoC on the left and the amount of domains scanned looking for a vulnerable software component on the right, where its vulnerability was confirmed by observing its software version.

\subsubsection{Takeaways}
The goal of this section was to identify features of the infrastructure hosting \pPDF s which may facilitate the upload of \pPDF s. 

Firstly, upon collecting ACL information for \empirical{\percBucketsObservedACL{}\%} of all active S3 buckets, we observed that all of them allowed unauthenticated users to perform operations, e.g., via the \texttt{FullControl} or the \texttt{Write} permission.
In the remaining cases, we found that most of the PDFs were offline or the buckets were non-existent by the time we visited them, which suggests the possibility of a prior cleanup action. Consistently with these observations, the buckets with observable IoCs counted \empirical{\urlsFromIoCBuckets{}} unique URLs leading to \pPDF s.

We crawled \empirical{\nDomainsCfInStudyNoWeebly{}} \webhSame{} and \uncat{} FQDNs successfully (e.g., no \texttt{Timeout} errors) and observed that \empirical{51\%} of them run outdated software components. Among them, the amount of domain suffering from Unrestricted File Upload is low (\empirical{2\%}), hinting at the fact that this might not be the primary mean used by attackers to upload \pPDF s.
In total, these domains served \empirical{\numprint{1075835}} \pPDF s.

Additionally, we confirmed that \empirical{16.4\%} of the \empirical{\nDomainsCfInStudyNoWeebly{}} websites were running at least one component of the ``CK'' family, facilitating file upload, serving \empirical{\numprint{190258}} \pPDF s.
We underline that this is a lower bound of the possible websites running these components, as we reduced the amount of website scanned due to the large daily amount of scans otherwise necessary. The number of IoCs observed on URL path hints at a higher number of websites, i.e., \empirical{21.3\%}.
Overall, our analyses observed indicators of compromise for \empirical{46\%} (\empirical{\urlsFromIoCUncatWebhsame{}}) of the URLs analyzed in $\S$ \ref{sec:security_indicators}.

\section{Use of Support Infrastructure}
\label{sec:campaigns-on-hosts}

Having identified the types of hosting most abused by cybercriminals and the solutions to upload \pPDF s on them, we proceed to measure the duration of this activity via the \pdfCheckPipeline{} module, answering \textbf{RQ 3}. 
These analyses are conducted on \pPDF{} links in the \dsFromUrl{}, having discarded those with an offline status.
Next, we group these PDFs by visual similarity using our \textit{Clustering Module} (see $\S$ \ref{sec:ML_detector}) and observe how these clusters distribute over the hosting types.

\subsection{Duration of Abuse}
\label{sec:duration-of-abuse}

\begin{figure}
	\centering
	\includegraphics[width=1\columnwidth]{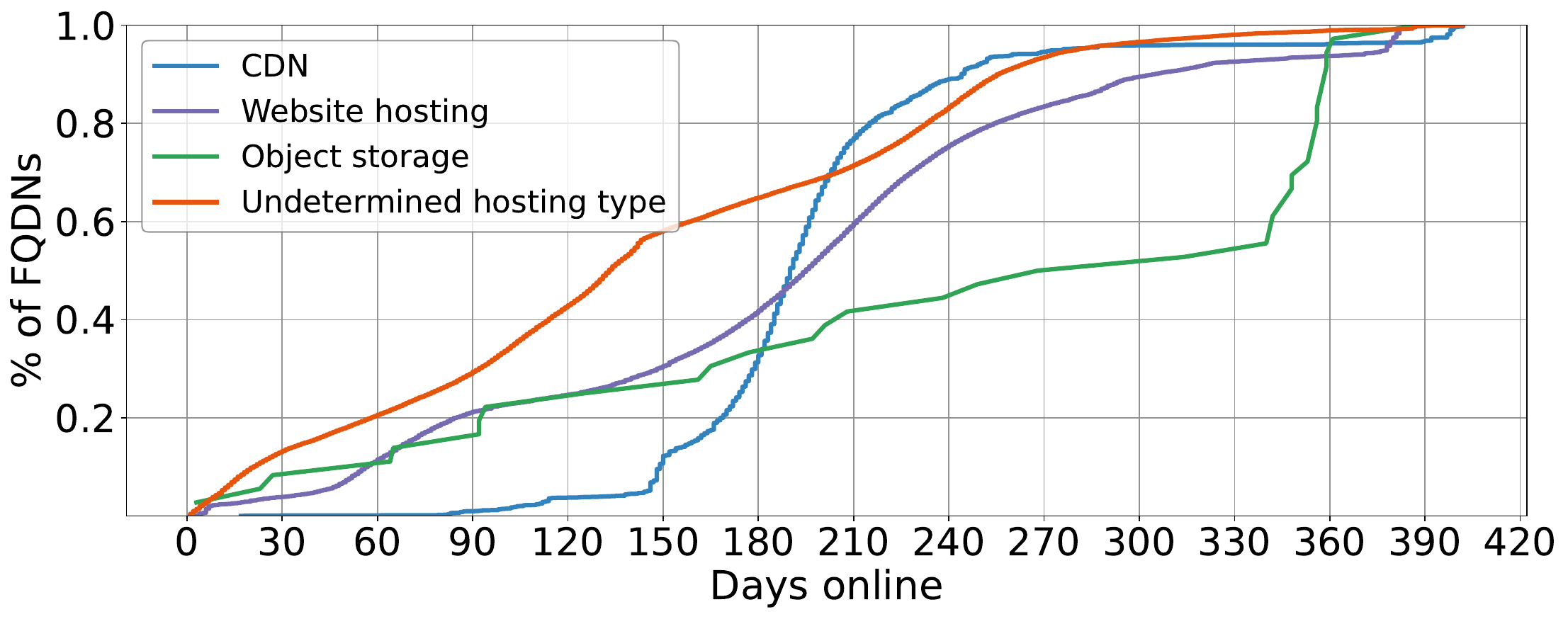}
	\caption{Distribution of \pPDF{} uptimes per hosting type, across our \empirical{\durationStudy{}}-month study.}
	\label{fig:uptime_days_abuse}
\end{figure}

We calculate the duration of the abuse as the mean uptime of each \pPDF{} hosted on a specific origin (with the granularity of a single day), as shown in \Cref{fig:uptime_days_abuse}.
Among the \empirical{\totHostingProviders{}} domain roots identified as hosting services, we observed the live abuse of \empirical{\provsObservedOnline{}} of them (the PDFs hosted on the remaining \empirical{\provsNotOnline{}} eTLD+1s were not online at the time we observed their URLs).

The average uptime for a single \pPDF{} is quite long, i.e., approximately \empirical{\avgPdfLifetimeMonths{}} months. However, due to the continuous upload of new PDFs on the same hosts, the overall abuse of hosting services extends even further, averaging around \empirical{\avgExtendedLifetimeMonths{}} months. It seems as if attackers persistently exploited these hosting services throughout our \empirical{\durationStudy{}} months of observations, with \empirical{\tldsMaxExtendedLifetime{}} domain roots receiving new uploads for this entire period.
The type of hosting providing the longest average PDF uptime is \empirical{\hostTypeLongestPDFLifetime{}}, where this value reaches \empirical{\avgHighestPdfLifetimeMonths{}} months.

\subsection{Distribution of PDF Clusters on Hosts}
\label{sec:campaigns_intro}
\begin{figure}
	\centering
	\includegraphics[width=0.8\columnwidth]{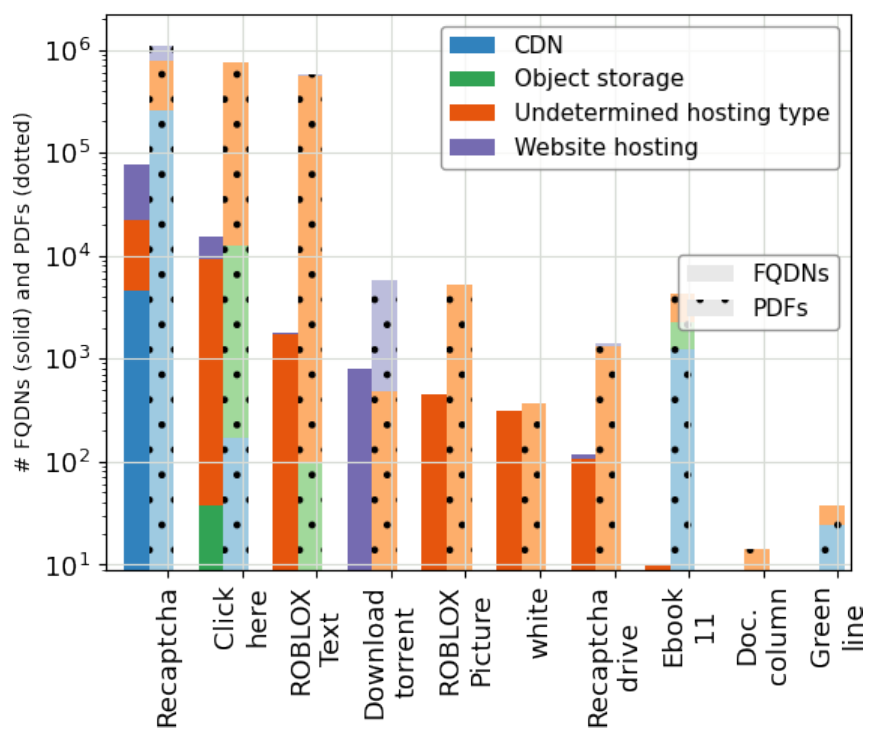}
	\caption{Stacked histogram showing clusters distribution across hosting types. Solid blocks represent the volume of FQDNs per cluster, while dotted blocks represent \pPDF{} volume.}
	\label{fig:distribution_camp_ovr_ht}
\end{figure}

In the \dsFromUrl{}, \pPDF{}s can be categorized, by visual bait similarity, into \empirical{\nCampaigns{}} groups, \empirical{seven} of which align with those previously reported in~\cite{stivala23from}, and four are newly identified. We observed fewer campaigns than \cite{stivala23from}, which could be attributed to attackers changing the visual baits used (since our data collection began \empirical{11} months after their experiments) or due to filtering out non-strictly-SEO campaigns. %
We gave the new clusters arbitrary names, i.e., \textit{Click here}, \textit{Doc. column}, \textit{Green line} and \textit{White}.

We use the group information to measure the distribution of live PDF clusters on different types of hosts, shown in \Cref{fig:distribution_camp_ovr_ht}.
When reading the graph by focusing on hosting types, we observe that all groups of \pPDF s make use of \uncat{} spaces, although to different extents.
\empirical{\campsOnlyUnkn{}} groups (\textit{Doc. column} and \textit{ROBLOX Picture}) upload PDFs solely on this category of hosts. \empirical{\campsUnknGtNinetyEight{}} more (\textit{ROBLOX Text} and \textit{White}) rely almost uniquely on these domains, hosting there more than \empirical{98\%} of their samples.
Conversely, PDFs belonging to the \textit{Download Torrent} group are uploaded almost exclusively \webhSame{} hosts (\empirical{\percTorrentPDFsUncat{}\%} of the samples belonging to this campaign).
Finally, we observe that Amazon's S3 storage is the type of hosting targeted by the highest number of clusters, as we could observe \empirical{six} different ones\footnote{\textit{Click here}, \textit{Ebook 11}, \textit{Recaptcha}, \textit{Recaptcha drive}, \textit{ROBLOX Text}, \textit{white}.}.

Conversely, when focusing on the PDF visual clusters, we observe that they differ in how they use hosting types. For example, the \textit{Ebook 11} cluster tends to perform large uploads of PDF on few hosts, as there large imbalance between the number of FQDNs where PDFs are uploaded and the number of uploaded PDFs (approximately \empirical{150} PDFs per eTLD+1, see \Cref{fig:distribution_camp_ovr_ht}). Differently, the \textit{Click-here} and \textit{Recaptcha} clusters distribute on average a smaller amount of PDFs per origin (approximately \empirical{2} per eTLD+1).
A third example is that of \textit{Download Torrent}, where there are large uploads of approximately \empirical{200} \pPDF s on \empirical{two} \uncat{} domains alongside smaller batches of uploads on many \webhSame{} FQDNs.

\subsection{Connection with IoCs}
We also observed that \empirical{\recaptchaWIoC{}\%} of \pPDF s belonging to the \textit{reCAPTCHA} campaign and \empirical{\robloxWIoC{}\%} of \pPDF s belonging to the \textit{Roblox Text} campaign are hosted on websites running one of the targeted plugins (regardless of their observed version).
These numbers represent a conservative estimate of the actual impact, as we chose to limit the number of IoCs tested to avoid excessive stress on the target websites.

\section{Fighting \PPDF s}
We now evaluate solutions to counter the distribution of \pPDF s, tackling \textbf{RQ 4}. We first consider existing solutions, in the form of blocklists, evaluating the protection they offer to users. Our observations indicate that blocklists provide limited user protections, motivating the need to take action against the spread of \pPDF s. 
Our proposed solution involves the notification of affected parties, where we report our observations on the presence of \pPDF s and on the status of the components running on the websites hosting the PDFs. 

\subsection{Blocklists}
\label{sec:anti-phishing-entities}
In this section, we investigate whether common blocklists, as VirusTotal (VT) and Google SafeBrowsing (GSB), take action against \pPDF s by blocklisting their URL. This would offer a viable protection to users, which would then be protected when accidentally visiting the page of the PDF.
We base our observations on \empirical{\lengthGSBDataCollection{}} months of Google SafeBrowsing and VirusTotal daily lookups (i.e., since the beginning of this study).
 
We request scan results for \empirical{4 thousand} \pPDF{} URLs daily to VirusTotal (approximately \empirical{50\%} of the daily amount) and receive a response in only \empirical{14\%} of the cases, where URLs are mostly flagged as malicious. This confirms the uncanny observation in~\cite{stivala23from} that URLs in \pPDF s are only partially scanned by VT and that this happens on the day the PDF is uploaded to the platform.
When considering the type of hosting, we observe that VirusTotal flags domains belonging to all four of them, with 
\empirical{\hostTypeMaxAvgScore{}} having the highest average rank (\empirical{\maxAvgPDFScore{}} AV engines) and \empirical{\hostTypeMinAvgScore{}} having the lowest average rank (\empirical{\minAvgPDFScore{}} AV engine).

Next, we observe that the number of URLs blocklisted by GSB is low, i.e., \empirical{\percBlockedOverSubm{}\%}.
These URLs belong to \empirical{\nBlocklGsbNetloc{}} domains, with a mean ratio of URLs per domain of \empirical{\avgBlocklGsbNetloc{}} (min \empirical{1}, max \empirical{1377}), which suggests that GSB is taking actions against \pPDF s and their hosts, blocklisting entire directories, but on a very small scale. Additionally, \empirical{99.7\%} of the blocklisted URLs belong to \uncat{} URLs, suggesting that GSB does not take any action against \pPDF s hosted on well-known, reputable domains.
When considering the overall lifetime of a \pPDF{}, as measured by the \pdfCheckPipeline{} module, we observe that a significant amount (\empirical{\percOnlineNowBlockl{}\%}) of the blocklisted PDFs is still online, which leads to think that blocklisting does not always correspond to a cleanup action.

\subsection{Vulnerability Notification} \pagesBudget{1.5}
\label{sec:vuln_notif}
Our next goal is to evaluate solutions beyond blocklisting to help reduce the spread of \pPDF s. 
One way to protect victims from the attack and, at the same time, to reduce the effectiveness of the SEO attack is taking down the PDFs by removing them from their location at the host.
We thus undertake a large-scale notification of the threat posed by \pPDF s to the affected parties.
Our primary goal is to observe the responsiveness of the hosting providers, measuring the amount of PDFs taken down as an effect of our reports.

\subsubsection{Setup of the Study}
We designed the notification procedure following best practices in this field~\cite{stock2016hey, stock2018didn, li2016you, soussi2020feasibility, Dietrich2018Investigating, kenneally2012menlo}. 
\paragraph{\textbf{Selection of Contacts}}
On Dec 1st, 2022 we select \empirical{\nPDFsNotified{}} \texttt{.pdf} links found online by our \textit{URL Analysis} module on the previous day and divide their FQDNs equally in Treatment and Control group (\empirical{\dimensionTestGrp{}} and \empirical{\dimensionControlGrp{}} respectively). Then, we look up their IP addresses and proceed to collect WHOIS records, obtaining \empirical{\totContactsForIPs{}} email contacts for \empirical{\totFetchedIPs{}} IPs. If necessary, we prioritize contacts from the same record, selecting \texttt{abuse@} contacts when present, \texttt{hostmaster@} contacts otherwise (following RFC 2142). If none of them are available, we choose one randomly.
We obtained no WHOIS record for \empirical{\totSynthDomains{}} domains, thus, we generate ``synthetic'' contacts by combining the aliases \texttt{abuse@}, \texttt{info@}, \texttt{security@}, \texttt{hostmaster@} with the domain name.

\paragraph{\textbf{Content and Timeline of Notification}} The notification e-mail briefly explains the threat posed by \pPDF s, then lists up to three \pPDF{} links among those hosted on up to three FQDNs belonging to the addressee. As a possible mitigation, we suggest the removal of the reported files and recommend a revision of the software components running at those domains.
A CSV attachment reports all \pPDF s links for all the domains belonging to the addressee. Finally, recipients are given the possibility to opt out of the study or reach back for any feedback. 
The full text of our notification message is reported in \Cref{app:notification_msg}.

Finally, we set a time window of 30 days, from Dec 1st to Dec 31st, 2022. We notified domains in the Treatment group once every ten days for a total of three times and notified the domains in the Control group at the end of the study. The choice for a ten-day time interval is motivated by the observations reported in~\cite{stock2018didn} where, in spite of the 14-day interval between each reminder, the number of fixes does not increase after ten days.

\paragraph{\textbf{Ethics}}
We did not seek IRB involvement for this procedure, addressing ethics concerns as follows.
Contact points (participants) were chosen depending on the presence of \pPDF s on their domains.
Participants were informed of the study and given the option to opt out immediately if in the Treatment group, or at the end of the monitoring period otherwise (Control group).
Although vulnerability notifications might represent an additional overhead for security operators at hosting providers, the benefit gained from \pPDF{} takedown and a security review of the software stack outweigh this cost. To reduce recipient overhead, we grouped domains per abuse contact. Finally, we did not collect any user data and sought to increase privacy of operators and providers by processing answers per anonymous ID rather than email address.

\subsubsection{Process}
A final amount of \empirical{\totSelectedContacts{}} contact emails was selected as recipient for the notification. The discrepancy between the number of contacts and FQDNs stems from them sharing the same eTLD+1 or a provider managing multiple FQDNs. Due to a technical problem, \empirical{\domainsNotReported{}} domains were not included in the reports or not reported at all, resulting in \empirical{\successfullySentFirstNot{}} emails being sent successfully.
These contacts were notified together with those in the Control group, but removed from the reports.

As part of the notification process, we excluded \empirical{one} contact, who asked to stop the analyses of the PDFs residing on their domain. Moreover, we adopted a ``cooperation policy'' whenever explicitly asked, e.g., we re-sent the attachment or provided clarification on the threat (\empirical{\nOurAnswers{}} replies), acknowledged false positives (\empirical{\nFalsePos{}} PDFs, \empirical{$<0.01\%$}), or submitted a copy of the report via a Web form (\empirical{\submittedForms{}} submissions).
Moreover, we estimated a lower bound of \empirical{\providersNeverReached{}} contacts we never reached by inspecting the headers of bounced emails.
As these providers could not be reached in the first round, we removed them from the Control group and did not notify them again.

\subsubsection{Effectiveness}

\begin{figure}
	\centering
	\includegraphics[width=1\linewidth]{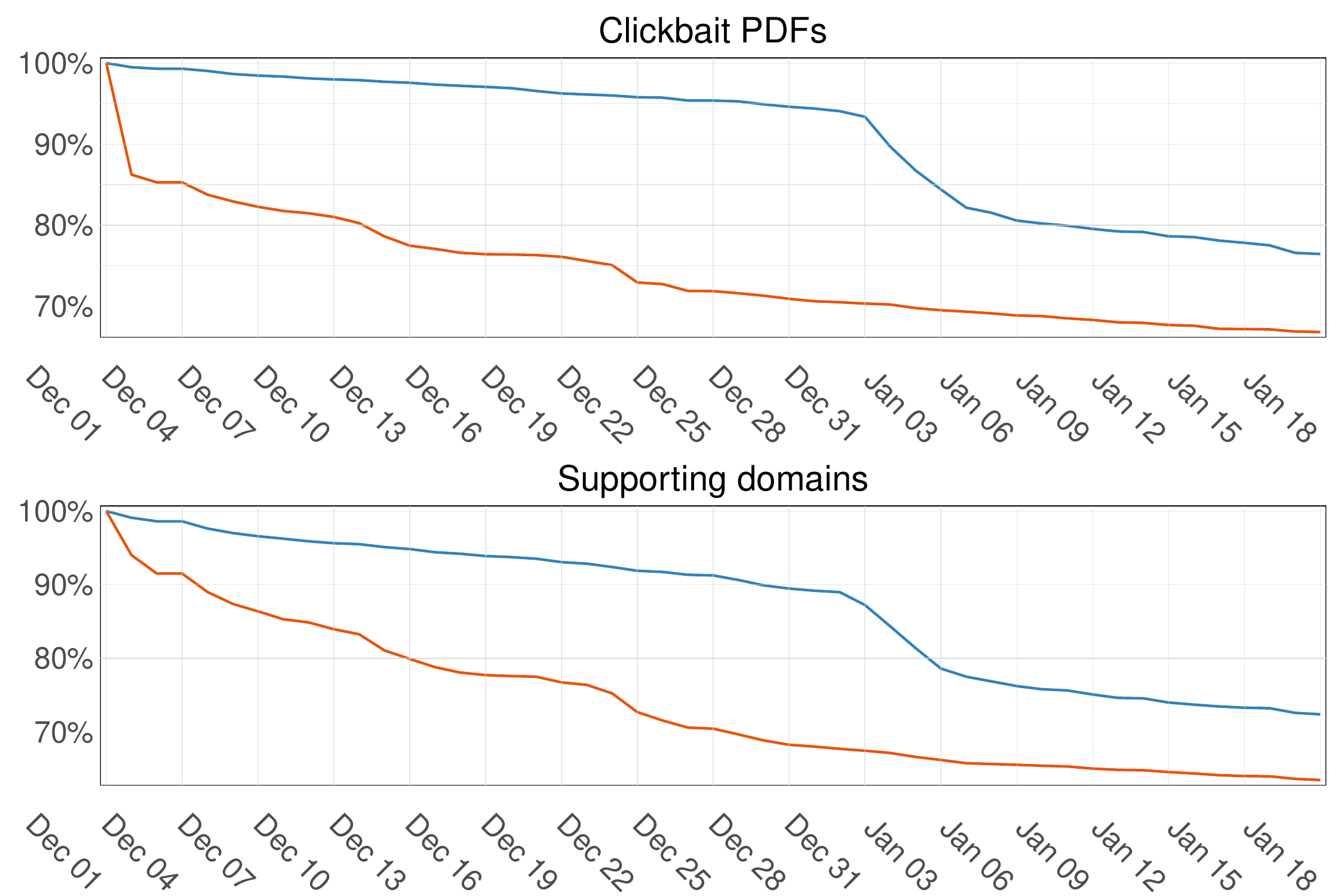}
	\caption{Takedown of \pPDF s and domains over time. The Treatment group is depicted in red and the Control group in blue.}
	\label{fig:vn_remediation}
\end{figure}

\Cref{fig:vn_remediation} shows the effectiveness of our notification by comparing the number of online \pPDF s in the Treatment and in the Control group.
The remediation rates are \empirical{29.567\%} for \pPDF s in the Treatment group and \empirical{6.055\%} for those in the Control group, where their difference is statistically significant with $\rho< .001$ (estimated by using a Generalized Linear Model~\cite{nelder1972generalized, mccullagh1989generalized}).
The number of online PDFs decreases sharply on the first days, while a less steep decrease is visible for the domains (\Cref{fig:vn_remediation}). One explanation for that may be that a few affected parties hosting a large number of \pPDF s took action immediately, while a larger number of entities, hosting less \pPDF s each, took longer to react.
The low-but-existent remediation rate for the Control group suggests the presence of some form of ``natural decay'', where a small fraction of \pPDF s go offline for causes not related to our notification. Nonetheless, the significantly higher remediation rate in the Treatment group shows an increased number of cleanup actions with respect to this phenomenon.

We observed that \empirical{no} affected party could remediate with respect to all reported domains (nor all PDFs, if on a single domain), and that \empirical{\percProvPartialFix{}\%} of the affected parties only partially remediated the notified issue.
In particular, \textit{(i)} \empirical{\provCleanEmBodyOverall{}} entities (\empirical{\percProvCleanEmBodyOverall{}\%}) only removed those PDFs listed in the email body, ignoring the attachment. \textit{(ii)} \empirical{\nProvsReExpl{}} entities (\empirical{\percProvsReExpl{}\%}) cleaned all or some of the reported PDFs. However, after the notification, we gained visibility into unseen (and unreported) \pPDF s hosted by them, which we observed stayed online. \textit{(iii)} \empirical{\nProvsDeepClean{}} entities (\empirical{\percProvsDeepClean{}\%}) performed a full cleanup and also removed any PDF observed after the notification.

We also monitored the presence of the reported PDFs on VirusTotal to observe if any of the affected parties submitted the PDFs as a result of our notification.
Given the large amount of \pPDF s involved and the limited API quota available to us, we opted to randomly sample unique PDFs, in equal amounts from the Treatment and the Control group. %
We fetched \empirical{\fhashFetchedVT{}} reports relative to notified \pPDF s.
\empirical{\fhashKnownVT{}} of these returned a record, where a negligible amount of them (\empirical{\percKnownAfterNotif{}\%}) was either first submitted or last seen after the start of the notification.
The number of domains hosting these PDFs belong almost equally to our Treatment and Control group.
Thus, it does not seem that submissions to VT were triggered by the notification.

\subsubsection{Long-Term Effectiveness}
\begin{figure}
	\centering
	\includegraphics[width=1\linewidth]{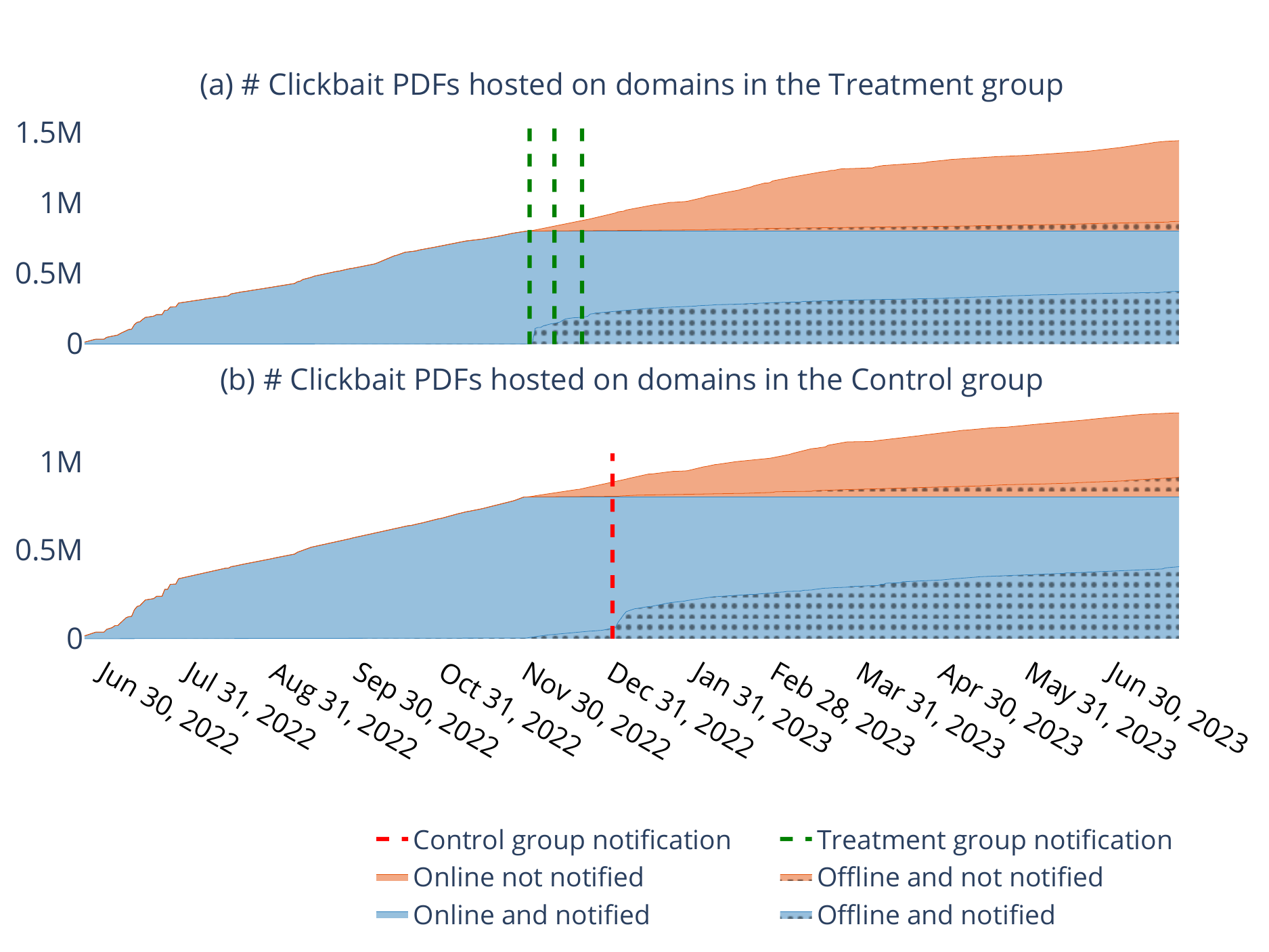}
	\caption{(a) Volume of PDFs in the Treatment group over time, online (solid color) and offline (dotted), versus new, unreported PDFs hosted by the same affected entities. (b) as for (a) but for PDFs in the Control group. (Control group volume is rescaled).}
	\label{fig:volume_new_VN_PDFs_over_time}
\end{figure}

We observed a moderate but positive response to the vulnerability notification in terms of PDFs that were cleaned up.
Our notification message clarified the possible presence of additional, unreported PDFs and recommended security audits on the software running on the affected domains.
We further investigated the long-time effects of our notification of the affected hosts, measuring how many of them still served \pPDF s, albeit unseen ones.
The observation of online unseen \pPDF s on notified hosts can be attributed to either new uploads from attackers or a partial cleanup by the responsible entity. \Cref{fig:volume_new_VN_PDFs_over_time} shows the online status of PDFs served by the domains involved in the notification. Starting from Dec. 1st, 2022 and Dec. 30th, 2022, we notice an increase in PDFs going offline, which mostly remains constant after the notification period concludes. Simultaneously, we continue to register unseen PDFs on the same origins, and their volume keeps growing over time. 
This disheartening finding shows that attackers have, and will continue to have, a relatively stable pool of hosts to upload PDFs in support of their attack.

Our findings also suggest that disclosing the presence of \pPDF s is a moderately effective means of reducing the volume of online PDFs at a specific point in time. However, it proves ineffective in enhancing the overall security level of the affected hosts.

\subsubsection{Feedback from Affected Parties}
We observed two main types of reactions to the notification, i.e., appreciation and interest versus an uncooperative attitude.

\paragraph{\textbf{Security Issues}}
A few affected parties confirmed our report and provided additional details or engaged in a conversation, allowing us to gain some invaluable insight on the issues they observed.
Five of them confirmed that their clients were running the plugins we identified in $\S$ \ref{sec:cksource_vulns}, specifically plugins of the ``CKFinder'' family or Formcraft for Wordpress.
Eight of them only generically replied that their client's CMS software was outdated (e.g., Joomla, Drupal) adding that they observed one or more PHP shells most likely used by the attackers to upload PDFs. One entity mentioned that their customer was running a custom web application.
Finally, in three cases, the answers reported that the website seemed to be abandoned by the customer, who was also unreachable.

\paragraph{\textbf{Not Phishing}}
Interestingly, one addressee answered all three notifications arguing that our report was unsubstantiated. They insisted that the reported PDFs did not pose any threat. Although we clarified the attack, they stated that they would not remove these legit files, as ``\textit{an interactive PDF with an attached hyperlink protected by recaptcha does not fall within the scope of phishing}'', referring to PDFs reproducing the reCAPTCHA service to trigger a click.

\section{Discussion} \pagesBudget{0.5}
\label{sec:limitations}

\textbf{\textit{SEO metric}.}
\label{sec:eval_seo_metric}
Our \textit{SEO metric} was designed with the goal of filtering out benign or not-\pPDF s to avoid processing personal data or poison our SEO-focused dataset. We perform a manual inspection to confirm that it only selects \pPDF s by inspecting up to \empirical{500} PDFs, fitting the SEO metric and randomly sampled from each cluster, for a total of \numprint{3000} PDFs. The manual analysis confirmed the null number of false positives. Conversely, some \pPDF s with too few backlinks may be excluded. In the worst-case scenario, where all PDFs failing the \textit{SEO metric} are clickbait PDFs, the false negatives would amount to \empirical{4.6\%} of 4.6 million links.
We inspected \empirical{\numprint{1000}} PDFs failing the \textit{SEO metric}, randomly sampled from the \dsVT{}, and observed a much lower amount of false negatives due to the presence of benign or non-\pPDF s.

\textbf{Development of \toolName{}.}%
The \pdfCheckPipeline{} module is a core component implementing the daily monitoring of online \pPDF s and enabling further analyses.
We ensured the reliability of its results by repeating requests to endpoints leading to an error three times, or by using a VPN service.
An interesting observation emerged where, in rare cases, an origin returned a different \texttt{HTTP} response for the same \texttt{.pdf} link. Specifically, we found that the \texttt{Content-Type} header differed between the \texttt{HEAD} request (not \texttt{application/pdf}) and the \texttt{GET} request (\texttt{application/pdf}). 
We examined a sample of 359 \texttt{.pdf} links marked as offline over the course of a week and did not observe any inconsistencies in the reported status.
Moreover, we observed \empirical{one} origin cloaking the content of the \texttt{HTTP} response, i.e., serving \pPDF s only when visited by a browser instance with enabled JavaScript, and \empirical{two} origins protected by the CloudFlare Bot Management service. \pdfCheckPipeline{} does not intend to bypass bot protections, and interestingly, we observed that such mechanisms are notably scarce in prevalence.

\textbf{Identification of Hosting Types.}
Our procedure for the identification of hosting types is based on observable metrics and indicators. All domain roots identified by our procedure correspond to an existing hosting service, confirming the validity of our methodology. 
We enriched this finding with the domain roots obtained from~\cite{similarweb}, which we verified belong to regional hosting providers.
We cannot rule out the possibility that attackers might also abuse other types of services to a lesser extent. For instance, \texttt{documentcloud.org}, a document sharing platform, served \empirical{121} \pPDF s at one point. However, we did not come across any further instances of such activity.

\textbf{Indicators of Compromise.}
Our findings show a grim picture of the landscape of software components running on the hosts part of the supporting infrastructure. Outdated and vulnerable components are especially present in \uncat{} origins, whose software stack is likely not managed by the service provider.
Ethical concerns on the traffic generated by our analyses on these origins limited the amount of scanning we performed to determine the component likely exploited by attackers to upload \pPDF s. Therefore, we believe the measurements we presented to be a lower bound of the amount of outdated or misconfigured software.

\textbf{Vulnerability Notification.}
Our vulnerability notification procedure effectively reduced the number of \pPDF s supporting the SEO attack and provided valuable insights into the software components running on a few notified websites, corroborating our automated analyses.

One methodological choice in this procedure may have influenced its outcome. Specifically, we formed the Treatment and Control groups based on domains instead of contact points. This decision aimed to achieve granularity in measuring remediation, focusing on individual PDF files rather than affected organizations or entities. However, we acknowledge that this approach might have increased the likelihood of cleanup for other websites in the Control group falling under the same entity's responsibility.
Moreover, our ``cooperation policy'', driven by a commitment to a safer Web, could have potentially influenced our results in a positive manner. We believe this impact to be limited, as we only engaged with \empirical{6\%} of the contact points. %

\textbf{External Threats to Validity.}
Our measurements might paint a less severe picture of the supporting infrastructure due to our partial visibility of the \pPDF{} threat. We mitigate this issue by collecting data from multiple sources: we build the \dsFromUrl{} starting from the \dsVT{} and observe that \empirical{\percNotSharedFromUrlPartners{}\%} of the total samples are not shared.
We believe that a complete picture of this ecosystem might be visible only to entities whose crawling and processing resources are far above ours.

\textbf{Looking Forward.}
\Cref{sec:vuln_notif} investigates the effectiveness of large-scale vulnerability notifications to address \pPDF s' abuse of hosting resources and protect users. While this approach proves effective in reducing online \pPDF s in the short term, there may be alternative methods to combat their distribution at various stages. For instance, making it more difficult for \pPDF s to rank high in search results could increase the attack cost and reduce the overall phenomenon. 
An implementation of this strategy could involve adding a module to a search engine crawler. The vast information available to search engines could serve as a crucial vantage point in preventing \pPDF s from achieving high rankings. 
Future research on \pPDF s could investigate which aspects of these documents are useful for detection.

\section{Ethical Considerations} \pagesBudget{0.7}
\label{sec:ethics}
We designed the experiments for this paper keeping a series of ethical concerns in mind. 
The daily scanning of online PDFs and indicators of compromise may raise ethical concerns. We followed established guidelines~\cite{durumeric2013zmap}, which included minimizing the frequency and load of experiments whenever possible (e.g., using \texttt{HEAD} requests instead of \texttt{GET}) and indicating the study's purpose, contact information, and opt-out option in the \texttt{User-Agent} header.
Additionally, we conduct a manual analysis to focus on high-probability IoC endpoints, minimizing unnecessary scanning, and follow best practices in vulnerability disclosure, refraining from testing endpoints where this is not allowed (avoid verifying vulnerable endpoints when this requires sending state-changing \texttt{POST} requests).
Finally, we reported all observed \pPDF s available with our large-scale vulnerability notification, started on Dec 1st, 2022. Our notification text explained about the threat posed by \pPDF s and included our contact points; we further gave participants the possibility to opt out of the study at any time.
We plan to conduct another notification campaign reporting the PDFs that are still online at submission time.

\section{Related Works}
We now review previous works connected to our study.

\textbf{Infrastructure Supporting Web Attacks.}
Previous studies have shown the use of abused infrastructure to support various attack campaigns, including malware delivery~\cite{nappa2013driving, nappa2014cyberprobe} and attack webpages~\cite{liao2016lurking, catakoglu2016automatic}. Nonetheless, these studies focus on the attack itself rather than studying how attackers use the support infrastructure, or they focus on a pre-determined list of hosting providers. 
Conversely, we uniquely investigate the supporting infrastructure without constraints on hosting service or provider. Our study shares similarities with Li et al.'s research~\cite{li2013finding}, both exploring malicious Web infrastructure, but differs in methodology, with Li et al. focusing on topological features of host interconnections.

\textbf{\PPDF s.}
Our study builds upon Stivala et al.'s work~\cite{stivala23from} which described the visual baits, structure, and distribution method of \pPDF s.
However, our approaches differ significantly in methodology and goals, as we examine attackers' use of supporting infrastructure, constructing an ad-hoc dataset, extracting information on (sub)domains and hosting types, and identifying vulnerable endpoints. 

\textbf{Vulnerability Indicators.} 
Previous works evaluated website security posture by detecting improper security headers and outdated software (e.g.,~\cite{goethem2014large}), WordPress plugins (e.g.,~\cite{kasturi2022mistrust}), or misconfigured S3 buckets (e.g.~\cite{continella2018there}).
Another line of work shows that search engines represent an alternative to Internet-wide scanning, as they can find indicators of vulnerable web servers~\cite{moore2009evil, john2010searching, zhang2012poisonamplifier, invernizzi2012evilseed}.
Part of our work touches the area of vulnerability scanning, as we verify the presence of exploitable vulnerabilities in hosts serving \pPDF s.
Specifically, we focus on identifying known security vulnerabilities or misconfigurations allowing file upload, which attackers might have relied on for the upload of \pPDF s. However, we do not aim to develop a general-purpose vulnerability scanner and limit our assessment to well-defined indicators.

\textbf{Abuse Monitoring in the Hosting Market.}
Finally, prior works examined abuse and cybercrime concentration from hosting providers' perspectives, particularly focusing on shared hosting and managed software components (e.g.,~\cite{tajalizadehkhoob2017herding, vasek2015hacking}).
They explored the correlation between an unsafe security posture at hosting providers and website compromise. Our work partially touches on services in the hosting market but does not target specific \textit{organizations} or providers and does not aim to establish statistically significant metrics linking \pPDF{} abuse to specific hosting services.

\section{Conclusion} \pagesBudget{0.33}
This paper presented a \durationEntireStudy{}-month study on the hosts supporting \pPDF{} attacks, counting \fqdnEntireStudy{} hosts and \totMonitoredLinksFromUrl{} links to \pPDF{}.
We observe that the websites supporting \pPDF{} attacks belong to different types of hosting, such as \objst{}, \webhDiff{} and \webhSame{} observed in our dataset, and that their continued abuse lasts \avgExtendedLifetimeMonths{} months on average.
Additionally, we developed hosting-type-specific analyses and identified six plugins and two web frameworks facilitating file upload \pPDF s files, and a large amount of websites running outdated software. 
Finally, we responsibly disclosed our findings via a large-scale vulnerability notification, observing a statistically significant decrease in the amount of online PDFs.
Nonetheless, we observed that most of the notified parties either suffered from re-uploads or performed partial cleanups, as their domains kept serving \pPDF s after the notification.
While a few parties took action against this threat, we observe that their impact is limited compared to the total volume of online PDFs and their hosting websites.

\bibliographystyle{plain}
\bibliography{bibliography}

\appendix
\section*{Appendices}
\subsection{The \toolName{} Pipeline}

\begin{table}
	\centering
	\footnotesize
	\begin{tabular}{c l r r l}
		\toprule
		ID                      &                 eTLD+1 &  \# FQDN &  \# URLs & Host. Type \\
		\midrule
		$\bullet$				&          amazonaws.com &       9 &   \numprint{49065} &      \objst{} \\
		$\bullet$				&      strikinglycdn.com &       1 &   \numprint{54052} &   \webhDiff{} \\
		$\bullet$				&           f-static.net &       1 &   \numprint{47931} &   \webhDiff{} \\
		$\bullet$				&                sqhk.co &       1 &   \numprint{15484} &   \webhDiff{} \\
		$\otimes$				&        squarespace.com &       1 &   \numprint{13000} &   \webhDiff{} \\
		$\bullet$				&            shopify.com &       1 &   \numprint{11994} &   \webhDiff{} \\
		$\bullet$				&    s123-cdn-static.com &       1 &   \numprint{10200} &   \webhDiff{} \\
		$\bullet$				&           filesusr.com &    \numprint{3241} &    \numprint{9829} &   \webhDiff{} \\
		$\bullet$				&           mozfiles.com &    \numprint{1741} &    \numprint{2366} &   \webhDiff{} \\
		\texttt{M}				&  s123-cdn-static-d.com &       1 &     531 &   \webhDiff{} \\
		\texttt{M}				&           s123-cdn.com &       1 &     139 &   \webhDiff{} \\
		\texttt{M}				&  s123-cdn-static-a.com &       1 &     138 &   \webhDiff{} \\
		\texttt{M}				&  s123-cdn-static-c.com &       1 &     134 &   \webhDiff{} \\
		\texttt{M}				&  s123-cdn-static-b.com &       1 &     124 &   \webhDiff{} \\
		$\bullet$				&             weebly.com &   \numprint{40803} &  \numprint{241092} &   \webhSame{} \\
		$\bullet$				&              epizy.com &    \numprint{4242} &    \numprint{5722} &   \webhSame{} \\
		$\bullet$				&            pbworks.com &    \numprint{1005} &    \numprint{4255} &   \webhSame{} \\
		$\bullet$				&          wordpress.com &    \numprint{1547} &    \numprint{4039} &   \webhSame{} \\
		$\bullet$				&                  rf.gd &    \numprint{3071} &    \numprint{3765} &   \webhSame{} \\
		$\bullet$				&           iblogger.org &    \numprint{1617} &    \numprint{1975} &   \webhSame{} \\
		$\bullet$				&              22web.org &    \numprint{1506} &    \numprint{1829} &   \webhSame{} \\
		$\otimes$				&              myhome.cx &      20 &    \numprint{1220} &   \webhSame{} \\
		$\bullet$				&       getenjoyment.net &     459 &    \numprint{1101} &   \webhSame{} \\
		$\bullet$				&     mywebcommunity.org &     419 &    \numprint{1059} &   \webhSame{} \\
		$\bullet$				&       myartsonline.com &     423 &    \numprint{1041} &   \webhSame{} \\
		$\bullet$				&      mypressonline.com &     432 &    \numprint{1040} &   \webhSame{} \\
		$\bullet$				&      onlinewebshop.net &     388 &     968 &   \webhSame{} \\
		$\bullet$				&      mygamesonline.org &     408 &     958 &   \webhSame{} \\
		$\bullet$				&     sportsontheweb.net &     406 &     951 &   \webhSame{} \\
		$\bullet$				&    scienceontheweb.net &     376 &     951 &   \webhSame{} \\
		$\bullet$				&    medianewsonline.com &     391 &     950 &   \webhSame{} \\
		$\bullet$				&         atwebpages.com &     390 &     940 &   \webhSame{} \\
		$\otimes$				&             linkpc.net &       4 &     896 &   \webhSame{} \\
		$\bullet$				&              66ghz.com &     171 &     260 &   \webhSame{} \\
		$\otimes$				&                 esy.es &       1 &     208 &   \webhSame{} \\
		$\otimes$				&           wpengine.com &       3 &     197 &   \webhSame{} \\
		$\otimes$				&       webhostmurah.com &       1 &     113 &   \webhSame{} \\
		$\otimes$				&         gridserver.com &       2 &      64 &   \webhSame{} \\
		$\otimes$				&                ovh.net &       2 &      45 &   \webhSame{} \\
		$\otimes$				&           yolasite.com &      44 &      44 &   \webhSame{} \\
		$\otimes$				&             hekko24.pl &       1 &      42 &   \webhSame{} \\
		$\otimes$				&          webbazaar.com &       2 &      35 &   \webhSame{} \\
		$\otimes$				&      000webhostapp.com &       1 &      34 &   \webhSame{} \\
		$\otimes$				&         pokladnicka.cz &       1 &      21 &   \webhSame{} \\
		$\otimes$				&         altervista.org &       1 &      21 &   \webhSame{} \\
		$\otimes$				&                 jpn.ph &       1 &      20 &   \webhSame{} \\
		$\otimes$				&              leszno.eu &       1 &      19 &   \webhSame{} \\
		$\otimes$				&             cafe24.com &       1 &      18 &   \webhSame{} \\
		$\otimes$				&              tenten.vn &       1 &      18 &   \webhSame{} \\
		$\otimes$				&       hostsolutions.ro &       1 &      15 &   \webhSame{} \\
		$\otimes$				&   belonnanotservice.ga &       1 &       8 &   \webhSame{} \\
		$\otimes$				&                home.pl &       1 &       1 &   \webhSame{} \\
		$\otimes$				&                webd.pl &       1 &       1 &   \webhSame{} \\
		$\otimes$				&           micron21.com &       1 &       1 &   \webhSame{} \\
		\bottomrule
	\end{tabular}
	\caption{Second-level domains and providers. Identification method (ID): $\bullet$ by threshold, \texttt{M} by manual analysis, $\otimes$ via Web analytics service~\cite{similarweb}.}
	\label{tab:IDed_providers}
\end{table}

\subsubsection{Clustering Module}
\label{sec:appendix-clustering}
Creating a clustering algorithm for the number and nature of samples presented a challenge, due to the necessity of handling shifting visual baits appearance and identifying new clusters without an available GPU. The pipeline operates in two steps: first, a CNN extracts a 32-dimensional feature vector from each sample, then, we use multiple iterations of DBSCAN to obtain document clusters. The embeddings are extracted daily, while the clustering procedure is manually triggered by a human operator.

\begin{table}
	\centering
	\footnotesize
	\begin{tabular}{l c c r r r c}
		\toprule
		Layer Name & Size In & Size Out & \# Kernel\\
		\midrule
		CNN-1 & $128 \times 128 \times 3 $ & $128 \times 128 \times 8 $ & $(3, 3)$ \\
		CNN-2 & $128 \times 128 \times 8 $ & $128 \times 128 \times 16$ & $(3, 3)$ \\
		CNN-3 & $128 \times 128 \times 16$ & $128 \times 128 \times 32$ & $(3, 3)$ \\
		MAXPOOL-1 & $128 \times 128 \times 32$ & $32 \times 32 \times 32$ & $(4, 4)$ \\
		CNN-4 & $32 \times 32 \times 32 $ & $32 \times 32 \times 64 $ &$ (3, 3)$ \\
		MAXPOOL-2 & $32 \times 32 \times 64 $&$ 8 \times 8 \times 64 $ & $(4, 4)$ \\
		CNN-5 & $8 \times 8 \times 64 $ & $8 \times 8 \times 128$ & $(3, 3)$ \\
		MAXPOOL-3 & $8 \times 8 \times 128$ & $2 \times 2 \times 128$  & $(4, 4)$ \\
		FLATTEN & $2 \times 2 \times 128$ & $512$ & \\
		FC & $512$ & $128$  &  \\
		FC & $128$ & $32$  &  \\
		L2 & $32$ & $32$ & \\
		\bottomrule
	\end{tabular}
	\caption{Details of the model architecture.}
	\label{tab:Model_architecture}
\end{table}

\textbf{The model.} The model takes the screenshot of the first page of the PDF as a $128\times128\times3$ matrix and returns a 32-dimensional vector. It consists of five convolutional blocks (a sequence of Convolution, BatchNormalization, PreLu, and dropout functions), three downsampling operations (MaxPooling), and two final FC layers. Additional details about the model are shown in \Cref{tab:Model_architecture}.
For training, we used a contrastive triplet loss with a margin of 0.2, implementing a semihard online triplet generation approach, as described in~\cite{schroff2015facenet}.

\textbf{Clustering.}
We use DBSCAN to group the PDF embeddings based on their appearance. To reduce the need for human intervention, we include a list of \empirical{20} pre-labeled items per group in the set of samples to be clustered. This list aids in automatically associating the clusters created by DBSCAN with existing known groups of visually-similar \pPDF s.
DBSCAN starts clustering samples with a default $\epsilon_0 = 0.25$.
If a computed cluster contains anchor samples from different campaigns, we reprocess its elements with $\epsilon_{i+1} = \epsilon_i - 0.01$ until the conflict is resolved. The clustering procedure is initiated manually by a human operator, who regularly inspects newly discovered clusters to verify the quality of the results.

\textbf{Validation.}
We manually inspected \numprint{3840} samples by selecting at most 500 random elements for each cluster.
In total, we found 105 misclassified samples, resulting in an error rate under 3\%.

\subsection{IoCs}
\subsubsection{Software Components Facilitating File Upload}
\label{app:details-exploits-cves}
This section presents the eight software components whose poor security status may have facilitated the upload of \pPDF s on a website.

\begin{table}
	\centering
	\footnotesize
	\begin{tabular}{llrr}
		\toprule
		SW Category &                  SW Name &  \# versions &  \# FQDNs \\
		\midrule
		CMS &             WordPress &      189 &    5912 \\
		CMS &                Joomla &        3 &     879 \\
		CMS &                Drupal &        3 &     347 \\
		Ecommerce &    Cart Functionality &        0 &    1828 \\
		Ecommerce &           WooCommerce &      159 &    1491 \\
		Ecommerce &  EasyDigitalDownloads &       11 &      24 \\
		Hosting panels &                 Plesk &        0 &    1029 \\
		Prog. language &                   PHP &      280 &   18279 \\
		Web servers &                Apache &       73 &   15065 \\
		Web servers &                 Nginx &       68 &    5592 \\
		Web servers &             LiteSpeed &        0 &    2013 \\
		WP plugins &        Contact Form 7 &       53 &    2105 \\
		WP plugins &             Yoast SEO &      196 &    1776 \\
		WP plugins &           WooCommerce &      159 &    1491 \\
		WP themes &                 Astra &       57 &     203 \\
		WP themes &       Hello Elementor &        9 &      87 \\
		WP themes &               OceanWP &       31 &      83 \\
		\bottomrule
	\end{tabular}
	\caption{Three most popular software components per category.}
	\label{tab:sw_components}
\end{table}

\textbf{FCKEditor, CKFinder, CKEditor, KCFinder.}
FCKEditor was a rich text editor first developed and released open-source by Frederico Caldeira Knabben in 2003~\cite{FCKEditorSource}. In January 2008, he released the first version of CKFinder~\cite{CKFinderOfficial}, ``the advanced file manager for FCKEditor''~\cite{CKFinderRelease}. 
FCKEditor has been assigned eight CVEs, among which \texttt{CVE-2006-2529}, affecting all versions until 2.3 Beta, allows an attacker to upload files of any type.
CKFinder has been assigned two CVEs, among which \texttt{CVE-2019-15862}, affecting all versions until  2.6.2.1, allows an attacker to upload files of any type.
In 2009, the author renames FCKEditor to CKEditor, releasing for the first time CKEditor 3~\cite{CKEditorRelease} and founding CKSource Holding LTD. The development of FCKEditor was discontinued.
Later, in 2015, right before the release of CKEditor 4.5, the plugin allegedly counted 15 million total downloads~\cite{CkEditor15M}.
CKEditor has been assigned \texttt{CVE-2015-9349} for a Cross-Site Scripting (XSS) vulnerability affecting all versions before 4.5.3.1. A popular exploit repository has shared the code to open a reverse shell in websites running CKEditor 4.4.7 or earlier~\cite{CKEditorSU}.

Finally, KCFinder was developed independently by Pavel Tzonkov~\cite{KCFinderOfficial} as a replacement to CKFinder, and to be compatible with FCKEditor and CKEditor. Its source code is still available~\cite{KCFinderSource}, although archived in 2021.
KCFinder has been assigned three CVEs, two of which are due to an XSS vulnerability and allow an attacker to inject and execute scripts. Affected are versions 3.20 and earlier, i.e., all versions. Multiple exploit repositories shared the code to exploit multiple vulnerabilities, e.g., Arbitrary File Upload in version 2.2~\cite{KCFinderAFU}, Shell Upload in version 2.53~\cite{KCFinderSU}.

\textbf{E-Learning Madrasah.}
This Web application was developed by the Indonesian Government as a response against the stop of all educational activities during the Covid-19 pandemic~\cite{sutiah2021software, madrasah_article}. Educational institutions (e.g., high schools) were equipped with an online platform (``E-learning Madrasah'') allowing all remote teaching activities.
This platform comes with the vulnerable component CKFinder installed, whose exploit code is publicly available~\cite{eLearnUFU}.

\textbf{Senayan Library Management System (SLiMS).}
This is an open-source web framework for library management developed in Jakarta. Its popularity might be higher in Indonesia, as all websites mounting this framework have a \texttt{.id} country code. Moreover, the manual analysis showed that most of these websites were websites of educational institutions.
SLiMS 7 and SLiMS 9 have been found vulnerable of multiple XSS, receiving two and three CVEs respectively, whose exploits are published in popular exploit databases~\cite{slims7SU, slims9SU}.

\textbf{FormCraft, Webform.}
FormCraft is a WordPress plugin offering form building functionalities~\cite{FormCraftOfficial}. Webform is a form builder plugin built for Drupal~\cite{WebformOfficial}.
FormCraft versions below 1.2.6 and below 3.6 have been assigned two CVEs for two XSS vulnerabilities, and a popular exploit repository published the code targeting FormCraft version 2.0 leading to Shell Upload~\cite{FormCraftSU}. Conversely, Webform was found vulnerable to multiple vulnerabilities, including an XSS introduced by the inclusion of the vulnerable CKEditor library~\cite{WebformXSS}.

\subsubsection{URL Path Indicators}
\label{app:details-path-indicators}

Below is the list of indicators of compromise, where the URL path segments give out the presence of a possibly vulnerable component.
\begin{itemize}
	\item SLiMS: keywords \texttt{\_\_statics}, \texttt{gudangsoal} or \texttt{repository} in the URL path.
	\item CkFinder: URL path, param, query or fragment contain the keywords \texttt{ckfinder} or \texttt{ckimage} or \texttt{kcfinder} or \texttt{ckeditor} or \texttt{fckeditor} in the URL path.
	\item Formcraft: keyword \texttt{formcraft} in the URL path.
	\item  WebForm: keyword \texttt{webform} in the URL path.
	\item  SuperForms: keyword \texttt{super-forms} in the URL path.
	\item  Formidable: keyword \texttt{formidable} in the URL path.
\end{itemize}

\subsection{Notification email}
\label{app:notification_msg}
I am a security researcher at \texttt{Institution Name} in \texttt{Country}. As part of an academic research project, we discovered that \texttt{N} of your domains (\texttt{domain1.com}, \texttt{domain2.com}, \texttt{domain3.com} among them) are used to host and distribute \texttt{M} \pPDF{} files. These files embed links leading visitors to malicious web pages delivering phishing attacks, malware, or online scams. Victims discover these \pPDF s with search engines such as Google and Bing, leveraging the reputation of your domains.

We do not know how exactly the attackers manage to upload these files in your domains and we believe that your domains may have a vulnerable or misconfigured component that enables unrestricted file uploads. Here is an example of three relative URLs to \pPDF s hosted by the above domains:

\begin{verbatim}
domain1:
/path/to/file/1.pdf
/path/to/file/2.pdf
/path/to/file/3.pdf
domain2.com:
/another/path/to/file/4.pdf
/another/path/to/file/5.pdf
/another/path/to/file/6.pdf
domain3.com:
/yet/another/path/7.pdf
/yet/another/path/8.pdf
/yet/another/path/9.pdf
\end{verbatim}

We attach a CSV listing the \pPDF s relative paths per domain. Please note that the list we provided might not be exhaustive as attackers may have uploaded new files after this notification.

MITIGATIONS:
As a first step, we encourage you to immediately remove these PDFs from your domains to hamper the effectiveness of the phishing campaign. However, we recommend a security review of your websites, looking for outdated, unpatched, vulnerable, or misconfigured software components to prevent attackers from uploading new files.

As part of our study, we will monitor the \texttt{N} domains to verify if they still serve such PDFs. You can opt out of this study by contacting us at \texttt{author email}. The details in this email should be sufficient for you to mitigate the problem, nonetheless, feel free to contact us at the same address should you have any question or feedback.

DISCLAIMER: This message is part of an academic research project. Researchers did not (and will not) attempt to reproduce the attack. We are not trying to sell any product or service, and we are not trying to obtain any bounty.

\end{document}